\documentclass[%
superscriptaddress,
 aps,
 twocolumn, prd
]{revtex4-1}

\usepackage{graphicx}
\usepackage{bm}% bold math
\usepackage[english]{babel}
\usepackage{csquotes}
\usepackage{amsmath}
\usepackage{amsthm}
\usepackage{latexsym}
\usepackage{amssymb}
\usepackage{mathrsfs}
\usepackage{epic}
\usepackage{epsfig}
\usepackage[dvipsnames]{xcolor}
\usepackage{dsfont}
\usepackage[colorlinks = true]{hyperref}
\usepackage{comment}
\usepackage[all]{xy}
\usepackage{enumitem}
\usepackage{eucal}
\usepackage{calc}
\usepackage[all]{xy}
\usepackage{epic}
\usepackage{tensor}
\usepackage{epsfig}
\usepackage[usenames,pdf]{pstricks}
\usepackage{pst-grad} % For gradients
\usepackage{pst-plot} % For axes
\usepackage[space]{grffile} % For spaces in paths
\usepackage{etoolbox} % For spaces in paths
\usepackage{subfigure}

\usepackage[final]{showkeys}

\newcommand{\R}{\mathbb{R}}

\newcommand{\hnabla}{\overset{h}{\nabla}}
\newcommand{\vnabla}{\overset{v}{\nabla}}

\numberwithin{equation}{section}
\renewcommand\theequation{\arabic{section}.\arabic{equation}}

\begin{document}

\title{Gravitational spin Hall effect of light}

	\author{Marius A. Oancea}
	\email{marius.oancea@aei.mpg.de}
	\address{Max Planck Institute for Gravitational Physics (Albert Einstein Institute), Am M\"uhlenberg 1, D-14476 Potsdam, Germany}
	
	\author{J\'er\'emie Joudioux}
	\email{jeremie.joudioux@aei.mpg.de}
	\address{Max Planck Institute for Gravitational Physics (Albert Einstein Institute), Am M\"uhlenberg 1, D-14476 Potsdam, Germany}
		 
	\author{I. Y. Dodin}
	\email{idodin@princeton.edu}
	\address{Department of Astrophysical Sciences, Princeton University, Princeton, New Jersey 08544, USA}
	    	
	\author{D. E. Ruiz}
	\email{deruiz@sandia.gov}
	\address{Sandia National Laboratories, P.O. Box 5800, Albuquerque, New Mexico 87185, USA}
		 
	\author{Claudio F. Paganini}
	\email{claudio.paganini@aei.mpg.de}
	\address{Fakult\"at f\"ur Mathematik, Universit\"at Regensburg, D-93040 Regensburg, Germany}
	\address{Max Planck Institute for Gravitational Physics (Albert Einstein Institute), Am M\"uhlenberg 1, D-14476 Potsdam, Germany}

	\author{Lars Andersson}
	\email{lars.andersson@aei.mpg.de}
	\address{Max Planck Institute for Gravitational Physics (Albert Einstein Institute), Am M\"uhlenberg 1, D-14476 Potsdam, Germany}

\begin{abstract}
The propagation of electromagnetic waves in vacuum is often described within the geometrical optics approximation, which predicts that wave rays follow null geodesics. However, this model is valid only in the limit of infinitely high frequencies. At large but finite frequencies, diffraction can still be negligible, but the ray dynamics becomes affected by the evolution of the wave polarization. Hence, rays can deviate from null geodesics, which is known as the gravitational spin Hall effect of light. In the literature, this effect has been calculated \textit{ad hoc} for a number of special cases, but no general description has been proposed. Here, we present a covariant Wentzel-Kramers-Brillouin analysis from first principles for the propagation of light in arbitrary curved spacetimes. We obtain polarization-dependent ray equations describing the gravitational spin Hall effect of light. We also present numerical examples of polarization-dependent ray dynamics in the Schwarzschild spacetime, and the magnitude of the effect is briefly discussed. The analysis reported here is analogous to that of the spin Hall effect of light in inhomogeneous media, which has been experimentally verified.
\end{abstract}

\maketitle

\section{Introduction}

The propagation of electromagnetic waves in curved spacetime is often described within the geometrical optics approximation, which applies in the limit of infinitely high frequencies \cite{MTW,gravitational_lenses_book}. In geometrical optics, Maxwell's equations are reduced to a set of ray equations and a set of transport equations along these rays. The ray equations are the null geodesics of the underlying spacetime, and the transport equations govern the evolution of the intensity and the polarization vector. In particular, the geometrical optics approximation predicts that the ray equations determine the evolution of the polarization vector and there is no backreaction from the polarization vector onto the ray equations. However, this model is valid only in the limit of infinitely high frequencies, and there has been interest in calculating the light propagation more accurately. At large but finite frequencies, diffraction can still be negligible but rays can deviate from geodesics. This is known as the gravitational spin Hall effect of light \cite{GSHE_review}.

The mechanism behind the spin Hall effect is the spin-orbit interaction \cite{SOI_review}, i.e., the coupling of the wave polarization (spin) with the translational (orbital) motion of the ray as a particle, resulting in polarization(spin)-dependent rays. Related phenomena are found in many areas of physics. In condensed matter physics, electrons traveling in certain materials experience a spin Hall effect, resulting in spin-dependent trajectories, and spin accumulation on the lateral sides of the material \cite{SHE_review1,SHE_review}. The effect was theoretically predicted by Dyakonov and Perel in 1971 \cite{originalSHE1, originalSHE2}, followed by experimental observation in 1984 \cite{originalSHE3} and 2004 \cite{originalSHE4}. In optics, the polarization-dependent deflection of light traveling in an inhomogeneous medium is known as the spin Hall effect of light \cite{SOI_review,SHEL_review}. The effect was predicted by several authors \cite{OpticalMagnus,SHE-L_original,SHE_original,Bliokh2004,Bliokh2004_1,Duval2006,Duval2007} and has recently been verified experimentally by Hosten and Kwiat \cite{SHEL_experiment} and also by Bliokh \textit{et~al.} \cite{Bliokh2008}. The spin Hall effect of light provides corrections to the geometrical optics limit, which scale roughly with the inverse of frequency. This, and several other effects, can be explained in terms of the Berry curvature \cite{Berry_CM1,SHE_QM2,Bliokh2008,SOI_review}.

There are several approaches aiming to describe the dynamics of spinning particles or wave packets in general relativity. Using a multipole expansion of the energy-momentum tensor, the dynamics of massive spinning test particles has been extensively studied in the form of the Mathisson-Papapetrou-Dixon equations \cite{mathisson2010republication,papapetrou1951spinning,tulczyjew1959motion,dixon1964covariant,dixon2015new}. A massless limit of these equations was derived by Souriau and Saturnini \cite{souriau1974modele,saturnini1976modele}, and particular examples adapted to certain spacetimes have been discussed in Refs.~\cite{Duval,Duval2018hzh, marsot}. Another commonly used method is the Wentzel-–Kramers–-Brillouin (WKB) approximation for various field equations on curved spacetimes. For massive fields, this has been done in Refs.~\cite{audretsch,rudiger} by considering a WKB approximation for the Dirac equation. For massless fields, using a WKB approximation for Maxwell's equations on a stationary spacetime, Frolov and Shoom derived polarization-dependent ray equations \cite{Frolov,Frolov2} (see also Refs.~\cite{covariantSpinoptics,spinorSpinoptics,spinorSpinoptics2,Harte2018}). With methods less familiar in general relativity, using the Foldy-Wouthuysen transformation for the Bargmann-Wigner equations in a perturbative way, Gosselin \textit{et~al.} derived ray equations for photons \cite{SHE_QM1} and electrons \cite{SHE_Dirac} traveling in static spacetimes (see also Refs.~\cite{Silenko2005,silenko2008foldy,obukhov2017general}). The gravitational spin Hall effect of gravitational waves was also considered in Refs.~\cite{covariantSpinoptics,SHE_GW}. However, as discussed in Ref.~\cite{GSHE_review}, there are inconsistencies between the predictions of these different models, and some of these models only work in particular spacetimes.

In this work, we are concerned with describing the propagation of electromagnetic waves in curved spacetime beyond the traditional geometrical optics approximation. We carry out a covariant WKB analysis of the vacuum Maxwell's equations, closely following the derivation of the spin Hall effect in optics \cite{Bliokh2008,Ruiz2015,Ruiz2017}, as well as the work of Littlejohn and Flynn \cite{Littlejohn1991}. As a result, we derive ray equations that contain polarization-dependent corrections to those of traditional geometrical optics and capture the gravitational spin Hall effect of light. As in optics, these corrections can be interpreted in terms of the Berry curvature. To illustrate the effect, we give some numerical examples of the effective ray trajectories in the Schwarzschild spacetime.

Our paper is organized as follows. In Sec.~\ref{sec:Maxwell_WKB}, we start by introducing the variational formulation of the vacuum Maxwell's equations. Then, we present the specific form of the WKB ansatz to be used, discuss the role of the Lorenz gauge condition, and state the assumptions that we are considering on the initial conditions. In Sec.~\ref{sec:Higher_GO}, we present the WKB approximation of the field action and the corresponding Euler-Lagrange equations. After analyzing these equations at each order in the geometrical optics parameter $\epsilon$, we obtain the well-known results of geometrical optics. The dynamics of the polarization vector is expressed in terms of the Berry phase. Finally, we derive an effective Hamilton-Jacobi system that contains $\mathcal{O}(\epsilon)$ corrections over the standard geometrical optics results. In Sec.~\ref{sec:effectivemotion}, we use the corrected Hamilton-Jacobi equation to derive the ray equations that account for the gravitational spin Hall effect of light. The gauge invariance of these equations is discussed, and noncanonical coordinates are introduced. In Sec.~\ref{sec:numerics}, we present some basic examples. For Minkowski spacetime, we analytically show how the effective ray equations reproduce the relativistic Hall effect \cite{Relativistic_Hall} and the Wigner translations of polarized electromagnetic wave packets \cite{Stone2015}. Using numerical computations, we consider the effective ray equations on a Schwarzschild background and compare with the results of Gosselin \textit{et~al}. \cite{SHE_QM1}. The magnitude of the effect is also estimated numerically. A summary of the main results, including the effective Hamiltonian and the effective ray equations, can be found in Sec.~\ref{sec:conclusions}.

\subsection*{Notations and conventions}
We consider an arbitrary smooth Lorentzian manifold $(M, g_{\mu \nu})$, where the metric tensor $g_{\mu \nu}$ has signature $-+++$. The absolute value of the metric determinant is denoted as $g = |\det g_{\mu \nu}|$. The phase space is defined as the cotangent bundle $T^*M$, and phase space points are denoted as $(x, p)$. The Einstein summation convention is assumed. Greek indices represent spacetime indices and run from $0$ to $3$. Latin indices from the beginning of the alphabet, $(a, b, c, \ldots)$, represent tetrad indices and run from $0$ to $3$. Latin indices from the middle of the alphabet, $(i, j, k, \ldots)$, label the components of $3$-vectors and run from $1$ to $3$. For the curvature, we use the conventions of Hawking and Ellis \cite{hawking1973}. Finally, we use the $\mathcal{O}$ notation as follows: a scalar function $f$ depending on a parameter $\epsilon$ satisfies $f(\epsilon) = \mathcal{O}(\epsilon^\alpha)$ if there is a constant $M$ such that $|f(\epsilon)| \leq M \epsilon^{\alpha}$ for small $\epsilon$.

\section{Maxwell's equations and the WKB approximation} \label{sec:Maxwell_WKB}

\subsection{Lagrangian formulation of Maxwell's equations}

Electromagnetic waves in vacuum can be described by the electromagnetic tensor $\mathcal{F}_{\alpha \beta}$. This is a skew-symmetric real $2$-form, which satisfies the vacuum Maxwell's equations \cite[Sec.~22.4]{MTW}
\begin{equation} \label{eq:Maxwell_F}
    \nabla{}^\alpha\mathcal{F}{}_{\alpha \beta} = 0, \qquad \nabla{}_{[\alpha}\mathcal{F}{}_{\beta \gamma]} = 0. 
\end{equation}
Solutions to Maxwell's equations can also be represented by introducing the electromagnetic four-potential $\mathcal{A}_\alpha$, which is a real $1$-form. Then, the electromagnetic tensor can be expressed as
\begin{equation}
\mathcal{F}{}_{\alpha \beta}  = 2\nabla{}_{[\alpha} \mathcal{A}{}_{\beta]},
\end{equation}
and Eq.~\eqref{eq:Maxwell_F} becomes \cite[Sec.~22.4]{MTW}
\begin{equation}
\label{eq:Maxwell_eq}
\hat{D}\indices{_\alpha^\beta} \mathcal{A}_{\beta}  = 0, \qquad \hat{D}\indices{_\alpha^\beta} = \nabla^\beta \nabla_\alpha - \delta^\beta_\alpha \nabla^\mu \nabla_\mu .
\end{equation}
This equation can be obtained as the Euler-Lagrange equation of the following action:
\begin{equation} \label{eq:Maxwell_action}
\begin{split}
    J &= \frac{1}{4} \int_M \mathrm{d}^4 x \, \sqrt{g} \, \mathcal{F}_{\alpha \beta} \mathcal{F}^{\alpha \beta} \\ &= \frac{1}{2} \int_M  \mathrm{d}^4 x \, \sqrt{g} \, \mathcal{A}^\alpha \hat{D}\indices{_\alpha^\beta} \mathcal{A}_\beta,
\end{split}
\end{equation}
where the last equality is obtained using integration by parts.

\subsection{WKB ansatz}\label{sec:wkb}

We assume that the vector potential admits a WKB expansion of the form
\begin{equation} \label{eq:WKB_Maxwell}
\begin{split}
    \mathcal{A}_\alpha (x) &= \mathrm{Re} \left[ A_\alpha(x, k(x), \epsilon) e^{i S(x) / \epsilon} \right], \\
    A_\alpha(x, k(x), \epsilon) &= {A_0}_\alpha(x, k(x)) + \epsilon {A_1}_\alpha(x, k(x)) + \mathcal{O}(\epsilon^2),
\end{split}
\end{equation}
where $S$ is a real scalar function, $A_\alpha$ is a complex amplitude, and $\epsilon$ is a small expansion parameter. The gradient of $S$ is denoted as
\begin{equation} \label{eq:grad_S}
    k_\mu(x) = \nabla_\mu S(x).
\end{equation}
Note that the ansatz \eqref{eq:WKB_Maxwell} differs from the classical WKB ansatz since the complex amplitude $A_\alpha$ depends on the phase gradient $k_\alpha(x)$. In other words, we assume that $A_\alpha$ is defined on the Lagrangian submanifold $x \mapsto (x, k(x)) \in T^*M$. Such a dependency can be found in standard textbooks, for example Ref.~\cite[Sec.~3.3]{MR1806388}. Up to an application of the chain rule, our Eq.~\eqref{eq:WKB_Maxwell} is equivalent to the standard WKB ansatz. In particular, the dependency of $A$ in $k$ appears naturally in the geometrical optics equation \eqref{eq:k_orth_A}, and we observe in Sec.~\ref{sec:polarization_basis} that the polarization vector and the polarization basis naturally depend on $k$, which is why the $k$ dependence was introduced in Eq.~\eqref{eq:WKB_Maxwell}.

The limit $\epsilon \ll 1$ indicates that the phase of the vector potential rapidly oscillates and its variations are much faster than those corresponding to the amplitude $A_\alpha(x,k,\epsilon)$. The role of the expansion parameter $\epsilon$ becomes clear if we consider a timelike observer traveling along the worldline $\lambda \mapsto y^\alpha (\lambda)$ with proper time $\lambda$. This observer measures the frequency 
\begin{equation}
    \omega = - \frac{ t^\alpha k_\alpha }{ \epsilon },
\end{equation}
where $t^\alpha = \textrm{d} y^\alpha / \textrm{d} \lambda$ is the velocity vector field of the observer. The phase function $S$ and $\epsilon$ are dimensionless quantities. In geometrized units, such that $c = G = 1$ \cite[Appendix F]{Wald_GR}, the velocity $t^\alpha $ is dimensionless, and $k_\alpha$ has dimension of inverse length. Hence, $\omega$ has the dimension of the inverse length, as expected for frequency. Then, the observer sees the frequency going to infinity as $\epsilon$ goes to $0$.

\subsection{Lorenz gauge }\label{sec:lorenzgauge}

In this section, we introduce the Lorenz gauge in the context of WKB approximations. We shall impose this gauge condition [cf. \eqref{eq:Lorenz_cond1} below] in the rest of the paper, with the exception of Sec.~\ref{sec:0thGO}, where it is relevant to discuss some aspects of geometrical optics without this condition. 

Maxwell's equations in the form \eqref{eq:Maxwell_eq} do not have a well-posed Cauchy problem. In particular, they admit pure gauge solutions. This problem is usually eliminated by introducing a gauge condition. Here we shall focus on the Lorenz gauge condition
\begin{equation} \label{eq:fullLorenz}
     \nabla_\alpha \mathcal{A}^\alpha  = 0.
\end{equation}
We reproduce here, in the context of a WKB analysis, the classical argument regarding the gauge fixing  for Maxwell's equations (see for instance \cite[Lemma 10.2]{MR2473363}). Using the identity
\begin{equation}
 \hat{D}\indices{_\alpha^\beta} \mathcal{A}_{\beta}  - \nabla_\alpha \nabla^{\beta} \mathcal{A}_{\beta}  = - \nabla^\beta \nabla_\beta \mathcal{A}_{\alpha} + R_{\alpha \beta} \mathcal{A}^\beta,
\end{equation}
one observes that, if Maxwell's equations \eqref{eq:Maxwell_eq} and the Lorenz gauge \eqref{eq:fullLorenz} are satisfied, then the wave equation 
\begin{equation}\label{eq:reduced}
  - \nabla^\beta \nabla_\beta \mathcal{A}_{\alpha} + R_{\alpha \beta} \mathcal{A}^\beta = 0
\end{equation}
holds. Conversely, by solving Eq.~\eqref{eq:reduced}, with Cauchy data satisfying constraint and gauge conditions, one obtains a solution to Maxwell's equations in the Lorenz gauge.

Note that we consider here approximate solutions to Maxwell's equations
\begin{equation}
\hat{D}\indices{_\alpha^\beta} \mathcal{A}_{\beta} =\mathcal{O}(\epsilon^0). 
\end{equation}
Hence, it is sufficient to consider that the Lorenz gauge is satisfied at the appropriate order:
\begin{equation} 
\nabla_\alpha \mathcal{A}^\alpha = \mathcal{O}(\epsilon^1).
\end{equation}
We reproduce the standard argument recovering Maxwell's equation in the Lorenz gauge from the wave Eq.~\eqref{eq:reduced}, taking into account that we are considering only approximate solutions. Assume that the wave equation holds:
\begin{equation} \label{eq:wave}
- \nabla^\beta \nabla_\beta \mathcal{A}_{\alpha} + R_{\alpha \beta} \mathcal{A}^\beta = \mathcal{O}(\epsilon^0).
\end{equation}
Upon inserting the WKB ansatz, this is equivalent to
\begin{equation}
\begin{split}
    k^\beta k_\beta {A_0}_{\alpha} &= 0, \\
    i k^\beta k_\beta {A_1}_{\alpha} + {A_0}_{\alpha} \nabla^{\beta} k_\beta + 2 k^\beta \nabla_{\beta} {A_0}_{\alpha} &= 0.
\end{split}
\end{equation}
Furthermore, assume that the initial data for the wave equation
\eqref{eq:wave} satisfy
\begin{equation} \label{eq:IDwave}
\begin{split}
k_\alpha {A_0}^{\alpha} &= 0, \\
\nabla_\alpha {A_0}^\alpha + i k_\alpha {A_1}^\alpha &= 0. \\ 
\end{split}
\end{equation}
Equation \eqref{eq:wave} implies that 
\begin{equation} \label{eq:gaugesource}
\nabla^\beta  \nabla_\beta \left( \nabla_\alpha \mathcal{A}^\alpha \right)=  \mathcal{O}(\epsilon^{-1}).
\end{equation}
The initial data \eqref{eq:IDwave} for Eq.~\eqref{eq:wave} imply that, initially,
\begin{equation} \label{eq:gaugesource1}
\nabla_\alpha \mathcal{A}^\alpha = \mathcal{O}(\epsilon^1).
\end{equation}
Observe that the condition 
\begin{equation} \label{eq:gaugesource2}
T^\beta \nabla_\beta \left( \nabla_\alpha \mathcal{A}^\alpha \right) = \mathcal{O}(\epsilon^0)
\end{equation}
is automatically satisfied, where $T^\beta$ is a unit future-oriented normal vector to the hypersurface on which initial data are prescribed.  Hence, the equation satisfied by the Lorenz gauge source function \eqref{eq:gaugesource} admits initial data as in Eqs.~\eqref{eq:gaugesource1} and \eqref{eq:gaugesource2} vanishing at the appropriate order in $\epsilon$ [at $\mathcal{O}(\epsilon^1)$ and $\mathcal{O}(\epsilon^0)$, respectively]. This implies that Maxwell's equations
\begin{equation}
\hat{D}\indices{_\alpha^\beta} \mathcal{A}{}_{\beta} = \mathcal{O}(\epsilon^0),
\end{equation}
which can be expanded as
\begin{equation}
\begin{split}
    k^\beta {A_0}_{[\beta}k_{\alpha]} = 0, \\
    2 k^\beta \nabla_\beta {A_0}_{\alpha} - \left( \nabla_\beta {A_0}^{\beta} + i k_\beta {A_1}^{\beta} \right) k_\alpha  - k^\beta \nabla_\alpha {A_0}_{\beta} \\
    - {A_0}^{\beta} \nabla_\beta k_\alpha  + {A_0}_{\alpha} \nabla_{\beta} k^{\beta}  + i k^\beta k_\beta {A_1}_{\alpha}=0,
\end{split}
\end{equation}
are satisfied in the Lorenz gauge
\begin{equation}\label{eq:Lorenz_cond1}
 \nabla_\alpha \mathcal{A}^\alpha  =  \mathcal{O}(\epsilon^{1})\Leftrightarrow  \left\{ \begin{array}{r}  k_\alpha {A_0}^{\alpha}   = 0 \\
\nabla_\alpha {A_0}^\alpha + i k_\alpha {A_1}^\alpha = 0 
  \end{array} \right..
\end{equation}

\subsection{Assumptions on the initial conditions} \label{sec:init_data}

We end Sec.~\ref{sec:Maxwell_WKB} by summarizing the initial conditions that we shall use in the WKB ansatz for Maxwell's equations. 

\begin{enumerate}
    \item The Lorenz gauge \eqref{eq:Lorenz_cond1} is satisfied initially. This condition is used to obtain a well-defined solution to the equations of motion, as discussed in Sec.~\ref{sec:lorenzgauge}.
    
    \item The initial phase gradient $k_\alpha$ is a future-oriented null covector. As will be seen, the condition that $k_\alpha$ is null is a compatibility condition that follows from the Euler-Lagrange equations and the Lorenz gauge condition \eqref{eq:Lorenz_cond1} at the lowest order in $\epsilon$; cf. dispersion relation \eqref{eq:HJ1_eps0} below.

    \item Initially, the beam has circular polarization; cf. Eq.~\eqref{eq:circpol}. In Sec.~\ref{sec:polarization_basis} we show that the initial state of circular polarization is conserved. In Sec.~\ref{sec:effective_HJ} this assumption ensures a consistent transition transition between the effective dispersion relation and the effective ray equations. Heuristically speaking, due to the spin Hall effect, a localized wave packet that initially has linear polarization can split into two localized wave packets of opposite circular polarization. While this does not represent a problem at the level of Maxwell's equations (which are partial differential equations), the same behavior cannot be captured by the effective ray equations (which are ordinary differential equations) obtained in Sec.~\ref{sec:effective_HJ}. 
\end{enumerate}

\section{Higher-order geometrical optics} \label{sec:Higher_GO}

\subsection{WKB approximation of the field action}

We compute the WKB approximation for our field theory by inserting the WKB ansatz \eqref{eq:WKB_Maxwell} in the field action \eqref{eq:Maxwell_action}:
\begin{equation} \label{eq:action_WKB1}
\begin{split}
    J &= \int_M  \mathrm{d}^4 x \, \sqrt{g} \, \mathrm{Re} \left( A^\alpha e^{i S/ \epsilon} \right) \hat{D}\indices{_\alpha^\beta} \mathrm{Re} \left( A_\beta e^{i S/ \epsilon} \right) \\
    &= \frac{1}{4} \int_M  \mathrm{d}^4 x \, \sqrt{g}\, \left[  A^{*\alpha} e^{-i S/ \epsilon} \hat{D}\indices{_\alpha^\beta} \left( A_\beta e^{i S/ \epsilon} \right) +\text{c.c.} \right] \\
    & \quad+ \frac{1}{4} \int_M  \mathrm{d}^4 x \, \sqrt{g}\, \left[  A^\alpha e^{i S/ \epsilon} \hat{D}\indices{_\alpha^\beta} \left( A_\beta e^{i S/ \epsilon} \right) +\text{c.c.} \right].
\end{split}
\end{equation}
If $S$ has a nonvanishing gradient, then $e^{i S/\epsilon}$ is rapidly oscillating. In this case, for $f$ sufficiently regular, the method of stationary phase \cite[Sec.~2.3]{duistermaat1996fourier} implies
\begin{equation}
    \int_M  \mathrm{d}^4 x \, \sqrt{g}\, e^{\pm i 2 S(x) / \epsilon} f(x) =  \mathcal{O}(\epsilon^2).
\end{equation}
Upon expanding the derivative terms in Eq.~\eqref{eq:action_WKB1}, and keeping only terms of the lowest two orders in $\epsilon$, we obtain the following WKB approximation of the field action [for convenience, we are shifting the powers of $\epsilon$, such that the lowest-order term is of $\mathcal{O}(\epsilon^0)$]:
\begin{equation} \label{eq:eff_action}
\begin{split}
    -\epsilon^2 J =& \int_{M}  \mathrm{d}^4 x \, \sqrt{g}  \Big[ D\indices{_\alpha^\beta} A^{*\alpha} A_\beta \\
    &- \frac{i \epsilon}{2} \vnabla{}^\mu D\indices{_\alpha^\beta} \left( A^{*\alpha} {\nabla}_\mu A_\beta -  A_\beta {\nabla}_\mu A^{*\alpha} \right) \Big] + \mathcal{O}(\epsilon^2),
\end{split}
\end{equation}
where
\begin{equation} \label{eq:vertical_D}
\begin{split}
    D\indices{_\alpha^\beta} &= \frac{1}{2}k_\mu k^\mu \delta_\alpha^\beta - \frac{1}{2} k_\alpha k^\beta, \\
    \vnabla{}^\mu D\indices{_\alpha^\beta} &= k^\mu \delta_\alpha^\beta - \frac{1}{2} \delta _\alpha^\mu k^\beta - \frac{1}{2} g^{\mu \beta} k_\alpha.
\end{split}  
\end{equation}
Here, $D\indices{_\alpha^\beta}$ represents the symbol \cite{Fulling1996} of the operator $\hat{D}\indices{_\alpha^\beta}$, evaluated at the phase space point $(x, p) = (x, k)$, and we are using the notation $\vnabla{}^\mu D\indices{_\alpha^\beta}$ for the vertical derivative (Appendix~\ref{app:derivative_TM}) of $D\indices{_\alpha^\beta}$, evaluated at the phase space point $(x, p) = (x, k)$.

The action depends on the following fields: $S(x)$, $\nabla_\mu S(x)$, $A_\alpha(x,\nabla S)$, $\nabla_\mu \left[ A_\alpha(x, \nabla S) \right]$, $A^{*\alpha}(x,\nabla S)$, $\nabla_\mu \left[ A^{*\alpha}(x, \nabla S) \right]$. Following the calculations in Appendix~\ref{app:ELE}, the Euler-Lagrange equations are
\begin{widetext}
\begin{align}
    D\indices{_\alpha^\beta} A_\beta - i \epsilon \left(\vnabla{}^\mu D\indices{_\alpha^\beta} \right) \nabla_\mu A_\beta - \frac{i \epsilon}{2} \left( \nabla_\mu \vnabla{}^\mu D\indices{_\alpha^\beta} \right) A_\beta &= \mathcal{O}(\epsilon^2), \label{eq:HJ1_fullA} \\
    D\indices{_\alpha^\beta} A^{*\alpha} + i \epsilon \left(\vnabla{}^\mu D\indices{_\alpha^\beta} \right) \nabla_\mu A^{*\alpha} + \frac{i \epsilon}{2} \left( \nabla_\mu \vnabla{}^\mu D\indices{_\alpha^\beta} \right) A^{*\alpha} &= \mathcal{O}(\epsilon^2), \label{eq:HJ2_fullA} \\
    \nabla_\mu \Bigg[ \left(\vnabla{}^\mu D\indices{_\alpha^\beta} \right) A^{*\alpha} A_\beta
    - \frac{i\epsilon}{2} \left( \vnabla{}^\mu \vnabla{}^\nu D\indices{_\alpha^\beta} \right) \left( A^{*\alpha} \nabla_\nu A_\beta - A_\beta \nabla_\nu A^{*\alpha} \right) \Bigg] &= \mathcal{O}(\epsilon^2) \label{eq:transp_fullA}.
\end{align}
\\

\end{widetext}
In the above equations, the symbol $D\indices{_\alpha^\beta}$ and its vertical derivatives are all evaluated at the phase space point $(x, k)$. Note that the same set of equations can be obtained in a more traditional way, by inserting the WKB ansatz \eqref{eq:WKB_Maxwell} directly into the field equation \eqref{eq:Maxwell_eq}, or by following the approach presented in Ref.~\cite{Dodin2018}. More generally, a detailed discussion about the variational formulation of the WKB approximation can be found in Ref.~\cite{tracy2014}.

\subsection{Zeroth-order geometrical optics}\label{sec:0thGO}

Starting with Eqs.~\eqref{eq:HJ1_fullA}-\eqref{eq:transp_fullA}, and keeping only terms of $\mathcal{O}(\epsilon^0)$, we obtain
\begin{align}
    D\indices{_\alpha^\beta} {A_0}_\beta &= 0, \label{eq:HJ1_eps0}\\
    D\indices{_\alpha^\beta} {A_0}^{*\alpha} &= 0, \label{eq:HJ2_eps0}\\
    \nabla_\mu \Bigg[ \left(\vnabla{}^\mu D\indices{_\alpha^\beta} \right) {A_0}^{*\alpha} {A_0}_\beta \Bigg] &= 0. \label{eq:transp_eps0}
\end{align}
Equation \eqref{eq:HJ1_eps0} can also be written as
\begin{equation} \label{eq:eigen_disp}
    D\indices{_\alpha^\beta} {A_0}_\beta  = \frac{1}{2} \left( k_\mu k^\mu \delta_\alpha^\beta -k_\alpha k^\beta \right) {A_0}_\beta = 0.
\end{equation}
The matrix $D\indices{_\alpha^\beta}$ admits two eigenvalues when $k_\alpha$ is not a null covector. The first eigenvalue is $\frac{1}{2} k_\mu k^\mu$ with eigenspace consisting of covectors perpendicular to $k_\alpha$. The second eigenvalue is $0$ with eigenvector $k_\alpha$. When $k_\alpha$ is null, the matrix $D\indices{_\alpha^\beta}$ is nilpotent. It admits a unique eigenvalue $0$ whose eigenspace is the orthogonal to $k_\alpha$, which contains the covector $k_\alpha$. 

The Lorenz gauge condition \eqref{eq:Lorenz_cond1} implies that ${A_0}_\alpha$ is orthogonal to $k_\alpha$. Hence, a necessary condition for Eq.~\eqref{eq:HJ1_eps0} to admit a nontrivial solution is that $k_\alpha$ is a null covector. It is also possible to deduce that $k_\alpha$ is a null covector without using the gauge condition. For completeness, we present this argument below.  

Equation \eqref{eq:eigen_disp} admits nontrivial solutions if and only if ${A_0}_\beta$ is an eigenvector of $D\indices{_\alpha^\beta}$ with zero eigenvalue. Two cases should be discussed: $k_\alpha$ is a null covector, or $k_\alpha$ is not a null covector. 

Assume first that $k_\alpha$ is not a null covector, $k^\mu k_\mu \neq 0$. Then, Eq.~\eqref{eq:eigen_disp} leads to
\begin{equation}
    \quad {A_0}_\alpha = \frac{ k^\beta {A_0}_\beta}{k_\mu k^\mu} k_\alpha.
\end{equation}
This entails that
\begin{equation}
{A_0}{}_{[\alpha} k{}_{\beta]} = 0 \quad \text{ or } \quad  \mathcal{F}{}_{\alpha \beta} =\nabla{}_{[\alpha } \mathcal{A}{}_{\beta]} = \mathcal{O}(\epsilon^0).
\end{equation}
In other words, when $k_\alpha$ is not a null covector, the corresponding solution is, at the lowest order in $\epsilon$, a pure gauge solution. Since the corresponding electromagnetic field vanishes, we do not consider this case further.

If $k_\alpha$ is null, $k^\mu k_\mu = 0$, Eq.~\eqref{eq:eigen_disp} implies
\begin{equation}
 k^\beta {A_0}_\beta = 0.
\end{equation}
This is consistent with the Lorenz gauge condition \eqref{eq:Lorenz_cond1} at the lowest order in $\epsilon$. A similar argument can be applied for the complex-conjugate Eq.~\eqref{eq:HJ2_eps0}, from which we obtain $k_\alpha {A_0}^{*\alpha} = 0$.

Using Eqs.~\eqref{eq:HJ1_eps0}-\eqref{eq:transp_eps0}, we obtain the well-known system of equations governing the geometrical optics approximation at the lowest order in $\epsilon$: 
\begin{align}
   k_\mu k^\mu & = 0, \label{eq:disp_0} \\ 
    k^\alpha {A_0}_\alpha  = k_\alpha {A_0}^{*\alpha} & = 0, \label{eq:k_orth_A}\\
 \nabla_\mu \left( k^\mu \mathcal{I}_0 \right)  & = 0,  \text{}  \label{eq:transp_0}
\end{align}
where $\mathcal{I}_0 = {A_0}^{*\alpha} {A_0}_\alpha$ is the lowest-order intensity (more precisely, $\mathcal{I}_0$ is proportional to the wave action density \cite{tracy2014}). Equation \eqref{eq:transp_0} is obtained from Eq.~\eqref{eq:transp_eps0} by using the orthogonality condition \eqref{eq:k_orth_A}. Using Eq.~\eqref{eq:grad_S}, we have
\begin{equation} \label{eq:sym_nabla_k}
    \nabla_\mu k_\alpha = \nabla_\alpha k_\mu,
\end{equation}
and we can use Eq.~\eqref{eq:disp_0} to derive the geodesic equation for $k_\mu$:
\begin{equation} \label{eq:geodesic}
    k^\nu \nabla_\nu k_\mu = 0.
\end{equation}

\subsection{First-order geometrical optics}

Here, we examine Eqs.~\eqref{eq:HJ1_fullA} and \eqref{eq:HJ2_fullA} at order $\epsilon^1$ only:
\begin{align}
    \begin{split}
    D\indices{_\alpha^\beta} {A_1}_\beta &- i \left(\vnabla{}^\mu D\indices{_\alpha^\beta} \right) \nabla_\mu {A_0}_\beta \\
    &\quad- \frac{i}{2} \left( \nabla_\mu \vnabla{}^\mu D\indices{_\alpha^\beta} \right) {A_0}_\beta = 0, \end{split}& \label{eq:transp_a0}\\
    \begin{split}
    D\indices{_\alpha^\beta} {A_1}^{*\alpha} &+ i \left(\vnabla{}^\mu D\indices{_\alpha^\beta} \right) \nabla_\mu {A_0}^{*\alpha} \\
    &\quad+ \frac{i}{2} \left( \nabla_\mu \vnabla{}^\mu D\indices{_\alpha^\beta} \right) {A_0}^{*\alpha} = 0. \end{split}& \label{eq:transp_a0*}
\end{align}
Using Eq.~\eqref{eq:vertical_D}, we can also rewrite Eq.~\eqref{eq:transp_a0} as follows:
\begin{equation} \label{eq:transp_a0_1}
\begin{split} 
    k^\mu \nabla_\mu {A_0}_\alpha &- \frac{1}{2} k_\alpha \nabla_\mu {A_0}^\mu - \frac{1}{2} k_\beta \nabla_\alpha {A_0}^\beta \\
    &- \frac{i}{2} k_\alpha k^\beta {A_1}_\beta + \frac{1}{2} {A_0}_\alpha \nabla_\mu k^\mu \\
    &- \frac{1}{4} {A_0}^\beta \nabla_\beta k_\alpha - \frac{1}{4} {A_0}^\beta \nabla_\alpha k_\beta = 0.
\end{split}
\end{equation}
Using Eq.~\eqref{eq:sym_nabla_k}, we can rewrite the last two terms as
\begin{equation}
    - \frac{1}{4} {A_0}^\beta \nabla_\beta k_\alpha - \frac{1}{4} {A_0}^\beta \nabla_\alpha k_\beta = -\frac{1}{2} {A_0}^\beta \nabla_\alpha k_\beta.
\end{equation}
Using Eq.~\eqref{eq:k_orth_A}, we also have
\begin{equation}
    0 = \nabla_\alpha \big( k_\beta {A_0}^\beta \big) = k_\beta \nabla_\alpha {A_0}^\beta + {A_0}^\beta \nabla_\alpha k_\beta.
\end{equation}
Then, Eq.~\eqref{eq:transp_a0_1} becomes
\begin{equation}
\begin{split}
    k^\mu \nabla_\mu {A_0}_\alpha &+ \frac{1}{2} {A_0}_\alpha \nabla_\mu k^\mu \\
    &- \frac{1}{2} k_\alpha \left( \nabla_\mu {A_0}^\mu + i k_\mu {A_1}^\mu \right) = 0.
\end{split}
\end{equation}
The last term can be eliminated by using the Lorenz gauge \eqref{eq:Lorenz_cond1}. The same steps can be applied to the complex-conjugate Eq.~\eqref{eq:transp_a0*}:
\begin{equation} \label{eq:transp_a0_2}
\begin{split}
    k^\mu \nabla_\mu {A_0}_\alpha + \frac{1}{2} {A_0}_\alpha \nabla_\mu k^\mu &= 0, \\
    k^\mu \nabla_\mu {A_0}^{*\beta} + \frac{1}{2} {A_0}^{*\beta} \nabla_\mu k^\mu &= 0.
\end{split}
\end{equation}
Furthermore, using the lowest-order intensity $\mathcal{I}_0$, we can write the amplitude in the following way:
\begin{equation}
    {A_0}_\alpha = \sqrt{\mathcal{I}_0} {a_0}_\alpha, \qquad    {A_0}^{*\alpha} = \sqrt{\mathcal{I}_0} {a_0}^{*\alpha},
\end{equation}
where ${a_0}_\alpha$ is a unit complex covector (i.e., ${a_0}^{*\alpha} {a_0}_\alpha = 1$) describing the polarization. Then, from Eq.~\eqref{eq:transp_a0_2}, together with Eq.~\eqref{eq:transp_0}, we obtain
\begin{equation}\label{eq:polarization}
    k^\mu \nabla_\mu {a_0}_\alpha = k^\mu \nabla_\mu {a_0}^{*\alpha} = 0.
\end{equation}
The parallel propagation of the complex covector ${a_0}_\alpha$ along the integral curve of $k^\mu$ is another well-known result of the geometrical optics approximation.

\subsection{The polarization vector in a null tetrad} \label{sec:polarization_basis}

We observed that the polarization vector satisfies the orthogonality condition
\begin{equation}
    k^\alpha {a_0}_{\alpha} = 0.
\end{equation}
Consider a null tetrad \cite[Sec.~3]{PenroseRindler1} $\{k_\alpha, n_\alpha, m_\alpha, \bar{m}_\alpha\}$ satisfying
\begin{equation}\label{eq:NPtetrad}
\begin{split}
    m_\alpha \bar{m}^\alpha = 1, \qquad k_\alpha n^\alpha = -1, \\
    k_\alpha k^\alpha = n_\alpha n^\alpha = m_\alpha m^\alpha = \bar{m}_\alpha \bar{m}^\alpha &= 0, \\
    k_\alpha m^\alpha = k_\alpha \bar{m}^\alpha = n_\alpha m^\alpha = n_\alpha \bar{m}^\alpha &= 0.
\end{split}
\end{equation}
Note that we use the metric signature opposite to that used in Ref.~\cite[Sec.~3]{PenroseRindler1}. The covectors $n_\alpha, m_\alpha, \bar{m}_\alpha$ are not assumed to be parallel-propagated along the geodesic generated by $k^\alpha$. It is only $k_\alpha$ that is parallel-propagated along the geodesic generated by $k^\alpha$, in accordance with Eq.~\eqref{eq:geodesic}.
Since the null tetrad is adapted to the covector $k_\alpha$, the orthogonality conditions \eqref{eq:NPtetrad} imply that $m_\alpha$ and $\bar{m}_\alpha$ are functions of $k_\alpha$. 
The polarization covector ${a_0}_\alpha$ is orthogonal to $k_\alpha$, so we can decompose it as
\begin{equation} \label{eq:a0_basis}
    {a_0}_\alpha(x, k) = z_1(x) m_\alpha(x, k) + z_2(x) \bar{m}_\alpha(x, k) + z_3(x) k_\alpha(x),
\end{equation}
where $z_1$, $z_2$, and $z_3$ are complex scalar functions. Since ${a_0}_\alpha$ is a unit complex covector, the scalar functions $z_1$ and  $z_2$ are  constrained by
\begin{equation}
z_1^* z_1 + z_2^* z_2 = 1.
\end{equation}
It is important to note that the decomposition \eqref{eq:a0_basis}, and more specifically, the choice of $m_\alpha$, requires choosing a null covector $n_\alpha$. Fixing $n_\alpha$ is equivalent to choosing a unit timelike covector field $t_\alpha$ that can represent a family of timelike observers. We can always take $n_\alpha$ as
\begin{equation} \label{eq:t2n}
     t_\alpha =  \dfrac{1}{2\epsilon \omega} k_\alpha + \epsilon \omega n_\alpha.
\end{equation}
Once $n_\alpha$ (or $t_\alpha$) is fixed, the remaining $\mathrm{SO}(2)$ gauge freedom in the choice of $m_\alpha$ is described by the spin rotation
\begin{equation} \label{eq:spin_rotation}
    k_\alpha \mapsto k_\alpha, \quad n_\alpha \mapsto n_\alpha, \quad m_\alpha \mapsto e^{i\phi(x) } m_\alpha,
\end{equation}
for $\phi(x)\in \R$. Polarization measurements will always depend on the choice of $m_\alpha$ and $\bar{m}_\alpha$. However, as shown in Sec.~\ref{sec: noncanonical}, the modified ray equations describing the gravitational spin Hall effect of light do not depend on the particular choice of $m_\alpha$ and $\bar{m}_\alpha$. Thus, we can work with any smooth choice of $m_\alpha$ and $\bar{m}_\alpha$ that satisfy Eq.~\eqref{eq:NPtetrad}.

Using Eqs.~\eqref{eq:a0_basis} and \eqref{eq:geodesic}, the parallel-transport equation for the polarization vector becomes
\begin{equation}
\begin{split}
    0 &= k^\mu \nabla_\mu {a_0}_\alpha \\
    &= z_1 k^\mu \nabla_\mu m_\alpha + z_2 k^\mu \nabla_\mu \bar{m}_\alpha + m_\alpha k^\mu \nabla_\mu z_1 \\
    &\qquad + \bar{m}_\alpha k^\mu \nabla_\mu z_2 + k_\alpha k^\mu \nabla_\mu z_3. 
\end{split}
\end{equation}
Contracting the above equation with $\bar{m}^\alpha$, $m^\alpha$, and $n^\alpha$, we obtain
\begin{equation} \label{eq:z_transport}
\begin{split}
    k^\mu \nabla_\mu z_1 &= -z_1 \bar{m}^\alpha k^\mu \nabla_\mu m_\alpha, \\
    k^\mu \nabla_\mu z_2 &= - z_2 m^\alpha k^\mu \nabla_\mu \bar{m}_\alpha, \\
    k^\mu \nabla_\mu z_3 &= - ( z_1 m^\alpha + z_2 \bar{m}^\alpha) k^\mu \nabla_\mu n_\alpha.
\end{split}
\end{equation}
Recall that in the above equations, the covectors $m_\alpha$ and $\bar{m}_\alpha$ are functions of $x$ and $k(x)$. The covariant derivatives are applied as follows:
\begin{equation}
\begin{split}
    k^\mu \nabla_\mu m_\alpha &= k^\mu \nabla_\mu \left[ m_\alpha (x, k) \right] \\
    &= k^\mu \left( \hnabla{}_\mu m_\alpha \right) (x, k)\\ &\qquad+ k^\mu \left( \nabla_\mu k_\nu \right) \left(\vnabla{}^\nu m_\alpha \right) (x, k) \\
    &= k^\mu \hnabla{}_\mu m_\alpha,
\end{split}
\end{equation}
where $\hnabla_\mu$ is the horizontal derivative (Appendix~\ref{app:derivative_TM}). It is convenient to introduce the two-dimensional unit complex vector
\begin{equation}
    z = \begin{pmatrix} z_1 \\ z_2 \end{pmatrix},
\end{equation}
which is analogous to the Jones vector in optics \cite{optics_book,Bliokh2009,Ruiz2015,Ruiz2017}. We also use the Hermitian transpose $z^\dagger$, defined as follows:
\begin{equation}
z^\dagger =  \begin{pmatrix} z_1^* & z_2^* \end{pmatrix}.
\end{equation}
Then, the equations for $z_1$ and $z_2$ can be written in a more compact form:
\begin{equation}
    k^\mu \nabla_\mu z = i k^\mu B_\mu \sigma_3 z,
\end{equation}
where $\sigma_3$ is the third Pauli matrix, 
\begin{equation}
    \sigma_3 = \begin{pmatrix} 1 & 0 \\ 0 & -1 \end{pmatrix},
\end{equation}
and $B_\mu$ is the real 1-form extending to general relativity the Berry connection used in optics \cite{Bliokh2009,Ruiz2015}:
\begin{equation} \label{eq:B_connection}
\begin{split}
    B_\mu(x, k) &= \frac{i}{2} \left( \bar{m}^{\alpha} \hnabla_\mu {m}_\alpha - {m}_\alpha \hnabla_\mu \bar{m}^{\alpha} \right) \\
    &= i \bar{m}^{\alpha} \hnabla_\mu {m}_\alpha.
\end{split}
\end{equation}
Furthermore, if we restrict $z$ to an affinely parametrized null geodesic $\tau \mapsto x^\mu(\tau)$, with $\dot{x}^\mu = k^\mu$, we can write
\begin{equation} \label{eq:z_dot}
    \dot z = i k^\mu B_\mu \sigma_3 z,
\end{equation}
where $\dot z = \dot x^\mu \nabla_\mu z$. Integrating along the worldline, we obtain
\begin{equation} \label{eq:z_integration}
    z(\tau) = \begin{pmatrix} e^{i \gamma(\tau)} & 0 \\ 0 & e^{-i \gamma(\tau)} \end{pmatrix} z(0),
\end{equation}
where $\gamma$ represents the Berry phase \cite{Bliokh2009,Ruiz2015},
\begin{equation}
    \gamma(\tau_1) = \int_{\tau_0}^{\tau_1} \mathrm{d} \tau k^\mu B_\mu.
\end{equation}
Using either Eq.~\eqref{eq:z_transport} or Eq.~\eqref{eq:z_dot}, we see that the evolution of $z_1$ and $z_2$ is decoupled in the circular polarization basis, and the following quantities are conserved along $k^\mu$:
\begin{equation}
\begin{split}
1 &= z_1^* z_1 + z_2^* z_2 = z^\dag z , \\
s &= z_1^* z_1 - z_2^* z_2 = z^\dagger \sigma_3 z.
\end{split}
\end{equation}
Based on our assumptions on the initial conditions (Sec.~\ref{sec:init_data}), we only consider beams which are circularly polarized, i.e. one of the conditions 
\begin{equation} \label{eq:circpol}
    z(0) =  \begin{pmatrix} 1 \\ 0 \end{pmatrix}   \qquad \text{or} \qquad z(0) =  \begin{pmatrix} 0 \\ 1 \end{pmatrix}
\end{equation}
holds. Thus, we have $s=\pm 1$, depending on the choice of the initial polarization state. 

The results described in this section are similar to the description of the polarization of electromagnetic waves traveling in a medium with an inhomogeneous index of refraction \cite{Bliokh2009}.

\subsection{Extended geometrical optics}

Now, we take Eqs.~\eqref{eq:HJ1_fullA}-\eqref{eq:transp_fullA}, but without splitting them order by order in $\epsilon$. Our aim is to derive an effective Hamilton-Jacobi system that would give us $\mathcal{O}(\epsilon)$ corrections to the ray equations.

\subsubsection{Effective dispersion relation}

By contracting Eq.~\eqref{eq:HJ1_fullA} with $A^{*\alpha}$ and Eq.~\eqref{eq:HJ2_fullA} with $A_\beta$, and also adding them together, we obtain the following equation:
\begin{equation} \label{eq:efective_disp_1}
\begin{split}
    &D\indices{_\alpha ^\beta} {A}^{*\alpha} {A}_\beta \\
    &- \frac{i \epsilon}{2} \left( \vnabla{}^\mu D\indices{_\alpha^\beta} \right) \left( {A}^{*\alpha} \nabla_\mu {A}_\beta - {A}_\beta \nabla_\mu {A}^{*\alpha} \right) = \mathcal{O}(\epsilon^2).
\end{split}
\end{equation}
Using Eqs.~\eqref{eq:vertical_D} and \eqref{eq:k_orth_A}, we can rewrite the above equation as follows:
\begin{equation}
\begin{split}
    \frac{1}{2}&k_\mu k^\mu ( {A_0}^{*\alpha} {A_0}_\alpha + \epsilon {A_0}^{*\alpha} {A_1}_\alpha + \epsilon {A_1}^{*\alpha} {A_0}_\alpha )\\
    &- \frac{i \epsilon}{2} k^\mu \left( {A_0}^{*\alpha} \nabla_\mu {A_0}_\alpha - {A_0}_\alpha \nabla_\mu {A_0}^{*\alpha} \right)\\
    &+ \frac{i \epsilon}{4} k_\alpha \left( {A_0}^{*\mu} \nabla_\mu {A_0}^\alpha - {A_0}^\mu \nabla_\mu {A_0}^{*\alpha} \right)  = \mathcal{O}(\epsilon^2).
\end{split}
\end{equation}
Using Eq.~\eqref{eq:k_orth_A}, we obtain
\begin{equation}
\begin{split}
    0 &= {A_0}^{*\mu} \nabla_\mu \left( k_\alpha {A_0}^{\alpha} \right) \\
    &= k_\alpha  {A_0}^{*\mu} \nabla_\mu {A_0}^\alpha + {A_0}^{*\mu}  {A_0}^\alpha \nabla_\mu k_\alpha,
\end{split}
\end{equation}
so we can write
\begin{equation}
\begin{split}
    \frac{i \epsilon}{4} k_\alpha &\left( {A_0}^{*\mu} \nabla_\mu {A_0}^\alpha - {A_0}^\mu \nabla_\mu {A_0}^{*\alpha} \right) \\
    &= - \frac{i \epsilon}{2}\nabla_\mu k_\alpha {A_0}^{*[\mu}  {A_0}^{\alpha]}\\
    &= 0,
\end{split}
\end{equation}
where the last equality is due to Eq.~\eqref{eq:sym_nabla_k}. Then, Eq.~\eqref{eq:efective_disp_1} becomes
\begin{equation} \label{eq:efective_disp_2}
\begin{split}
    \frac{1}{2}& k_\mu k^\mu ( {A_0}^{*\alpha} {A_0}_\alpha + \epsilon {A_0}^{*\alpha} {A_1}_\alpha + \epsilon {A_1}^{*\alpha} {A_0}_\alpha ) \\
    &- \frac{i \epsilon}{2} k^\mu \left( {A_0}^{*\alpha} \nabla_\mu {A_0}_\alpha - {A_0}_\alpha \nabla_\mu {A_0}^{*\alpha} \right) = \mathcal{O}(\epsilon^2).
\end{split}
\end{equation}
Let us introduce the $\mathcal{O}(\epsilon^1)$ intensity
\begin{equation} \label{eq:def_I}
\begin{split}
\mathcal{I} &= \mathcal{A}^{*\alpha} \mathcal{A}_\alpha \\
            &= {A_0}^{*\alpha} {A_0}_\alpha + \epsilon {A_0}^{*\alpha} {A_1}_\alpha + \epsilon {A_1}^{*\alpha} {A_0}_\alpha + \mathcal{O}(\epsilon^2).
\end{split}
\end{equation}
Then, we can rewrite the amplitude as
\begin{equation}
    A_\alpha = \sqrt{\mathcal{I}} a_\alpha = \sqrt{\mathcal{I}}\left( {a_0}_\alpha + \epsilon {a_1}_\alpha  \right) +\mathcal{O}(\epsilon^2),
\end{equation} 
where $a_\alpha$ is a unit complex covector. Then, from Eq.~\eqref{eq:efective_disp_2} we obtain
\begin{equation}
    \frac{1}{2}k_\mu k^\mu - \frac{i\epsilon}{2} k^\mu \left( {a_0}^{*\alpha} \nabla_\mu {a_0}_\alpha - {a_0}_\alpha \nabla_\mu {a_0}^{*\alpha} \right) = \mathcal{O}(\epsilon^2).
\end{equation}
This can be viewed as an effective dispersion relation, containing $\mathcal{O}(\epsilon)$ corrections to the geometrical optics equation \eqref{eq:disp_0}. Finally, let us introduce
\begin{equation}
    K_\mu = k_\mu - \frac{i\epsilon}{2} \left( {a_0}^{*\alpha} \nabla_\mu {a_0}_\alpha - {a_0}_\alpha \nabla_\mu {a_0}^{*\alpha} \right)
\end{equation}
and rewrite the effective dispersion relation as
\begin{equation} \label{eq:eff_disp}
    \frac{1}{2} K_\mu K^\mu  = \mathcal{O}(\epsilon^2).
\end{equation}
It is worth noting that this equation can also be obtained directly from the effective field action \eqref{eq:eff_action}, specifically by varying the latter with respect to $\mathcal{I}$.

\subsubsection{Effective transport equation}

Using Eqs.~\eqref{eq:vertical_D}, \eqref{eq:disp_0}, and \eqref{eq:k_orth_A}, the effective transport equation \eqref{eq:transp_fullA} becomes
\begin{equation}
\begin{split}
    \nabla_\mu \Bigg[& k^\mu \left( {A_0}^{*\alpha} {A_0}_\alpha + \epsilon {A_0}^{*\alpha} {A_1}_\alpha + \epsilon {A_1}^{*\alpha} {A_0}_\alpha \right) \\
    &- \frac{i\epsilon}{2} g^{\mu \nu} \left( {A_0}^{*\alpha} \nabla_\nu {A_0}_\alpha - {A_0}_\alpha \nabla_\nu {A_0}^{*\alpha} \right) \\
    &+ \frac{i \epsilon}{4} \left( {A_0}^{*\alpha} \nabla_\alpha {A_0}^\mu - {A_0}^\mu \nabla_\alpha {A_0}^{*\alpha} \right) \\
    &+ \frac{i \epsilon}{4} \left( {A_0}^{*\mu} \nabla_\alpha {A_0}^\alpha - {A_0}^\alpha \nabla_\alpha {A_0}^{*\mu} \right) \\
    &-\frac{\epsilon}{2} k_\alpha \left( {A_0}^{*\mu} {A_1}^\alpha + {A_1}^{*\alpha} {A_0}^\mu \right) \Bigg] = \mathcal{O}(\epsilon^2).
\end{split}
\end{equation}
We can perform the following replacements in the above equation:
\begin{equation}
    \begin{split}
        {A_0}^{*\alpha} \nabla_\alpha {A_0}^\mu = \nabla_\alpha \left( {A_0}^{*\alpha} {A_0}^\mu \right) - \nabla_\alpha {A_0}^{*\alpha} {A_0}^\mu, \\
        \nabla_\alpha {A_0}^{*\mu} {A_0}^\alpha = \nabla_\alpha \left( {A_0}^{*\mu} {A_0}^\alpha \right) - {A_0}^{*\mu} \nabla_\alpha {A_0}^\alpha.
    \end{split}
\end{equation}
After rearranging terms, the effective transport equation becomes
\begin{equation}
\begin{split}
    \nabla_\mu \Bigg[& k^\mu \left( {A_0}^{*\alpha} {A_0}_\alpha + \epsilon {A_0}^{*\alpha} {A_1}_\alpha + \epsilon {A_1}^{*\alpha} {A_0}_\alpha \right) \\
    &- \frac{i\epsilon}{2} g^{\mu \nu} \left( {A_0}^{*\alpha} \nabla_\nu {A_0}_\alpha - {A_0}_\alpha \nabla_\nu {A_0}^{*\alpha} \right) \\
    &- \frac{i \epsilon}{2} {A_0}^{\mu} \left( \nabla_\alpha {A_0}^{*\alpha} - i k_\alpha {A_1}^{*\alpha} \right) \\
    &+ \frac{i \epsilon}{2} {A_0}^{*\mu} \left( \nabla_\alpha {A_0}^\alpha + i k_\alpha {A_1}^\alpha \right) \\
    &+ \frac{i \epsilon}{4} \nabla_\alpha \left( {A_0}^{*[\alpha} {A_0}^{\mu]} \right) \Bigg] = \mathcal{O}(\epsilon^2).
\end{split}
\end{equation}
The last term above vanishes due to the symmetry of the Ricci tensor:
\begin{equation}
\begin{split}
    \nabla_\mu \nabla_\alpha \left( {A_0}^{*[\alpha} {A_0}^{\mu]} \right) &= \nabla_{[\mu} \nabla_{\alpha]} \left( {A_0}^{*\alpha} {A_0}^{\mu} \right)\\
    &= \left(R\indices{_{\alpha \nu \mu}^\nu} -  R\indices{_{\mu \nu \alpha}^\nu} \right) {A_0}^{*\alpha} {A_0}^{\mu} \\
    &= \left(R_{\alpha \mu} -  R_{\mu \alpha} \right) {A_0}^{*\alpha} {A_0}^{\mu} \\
    &= 0.
\end{split}
\end{equation}
Furthermore, after using the Lorenz gauge condition \eqref{eq:Lorenz_cond1}, we are left with the following form of the effective transport equation:
\begin{equation}
\begin{split}
    &\nabla_\mu \Bigg[ k^\mu ( {A_0}^{*\alpha} {A_0}_\alpha + \epsilon {A_0}^{*\alpha} {A_1}_\alpha + \epsilon {A_1}^{*\alpha} {A_0}_\alpha ) \\
    &- \frac{i\epsilon}{2} g^{\mu \nu} \left( {A_0}^{*\alpha} \nabla_\nu {A_0}_\alpha - {A_0}_\alpha \nabla_\nu {A_0}^{*\alpha} \right) \Bigg] = \mathcal{O}(\epsilon^2).
\end{split}
\end{equation}
Introducing the intensity $\mathcal{I}$ and the vector $K^\mu$, we obtain
\begin{equation} \label{eq:eff_transp}
\begin{split}
    \nabla_\mu \Bigg\{ \mathcal{I} &\left[ k^\mu - \frac{i\epsilon}{2} g^{\mu \nu} \left( {a_0}^{*\alpha} \nabla_\nu {a_0}_\alpha - {a_0}_\alpha \nabla_\nu {a_0}^{*\alpha} \right) \right] \Bigg\} \\
    &= \nabla_\mu \left( \mathcal{I} K^\mu \right) = \mathcal{O}(\epsilon^2).
\end{split}
\end{equation}
This is an effective transport equation for the intensity $\mathcal{I}$, which includes $\mathcal{O}(\epsilon)$ corrections to the geometrical optics Eq.~\eqref{eq:transp_0}. As discussed in Ref.~\cite{tracy2014}, the direction of $K^\mu$ coincides with the direction of the wave action flux.

\section{Effective ray equations}\label{sec:effectivemotion}

\subsection{Hamilton-Jacobi system at the leading order}

The lowest-order geometrical optics equations \eqref{eq:disp_0} and \eqref{eq:transp_0} can be viewed as a system of coupled partial differential equations:
\begin{align}
   \frac{1}{2}g^{\mu \nu} k_\mu k_\nu &= 0, \label{eq:HJ_eq_0}\\
    \nabla_\mu \left( \mathcal{I}_0 k^\mu \right) &= 0, \label{eq:transp_HJ_0}
\end{align}
where $k_\mu = \nabla_\mu S$. Equation \eqref{eq:HJ_eq_0} is a Hamilton-Jacobi equation for the phase function $S$, and Eq.~\eqref{eq:transp_HJ_0} is a transport equation for the intensity $\mathcal{I}_0$ \cite{HJ_transport}. The Hamilton-Jacobi equation can be solved using the method of characteristics. This is done by defining a Hamiltonian function on $T^*M$, such that
\begin{equation} \label{eq:HJ_0}
    H \left(x, \nabla S \right) = \frac{1}{2}g^{\mu \nu} k_\mu k_\nu = 0.
\end{equation}
It is obvious that in this case, the Hamiltonian is
\begin{equation} \label{eq:H_0}
    H(x, p) = \frac{1}{2}g^{\mu \nu} p_\mu p_\nu.
\end{equation}
Note that in contrast to the dispersion relation \eqref{eq:HJ_0}, the Hamiltonian \eqref{eq:H_0} is a function on the whole phase space $T^*M$, with $p_\mu$ being an arbitrary covector. Hamilton's equations take the following form:
\begin{align} 
    \dot{x}^\mu &= \frac{\partial H}{\partial p_\mu} = g^{\mu \nu} p_\nu, \label{eq:EOM_0_x}\\
    \dot{p}_\mu &= -\frac{\partial H}{\partial x^\mu} = -\frac{1}{2} \partial_\mu g^{\alpha \beta} p_\alpha p_\beta. \label{eq:EOM_0_p}
\end{align}
Given a solution $\{x^\mu(\tau), p_\mu(\tau)\}$ for Hamilton's equations, we obtain a solution of the Hamilton-Jacobi Eq.~\eqref{eq:HJ_0} by taking \cite[p. 433]{goldstein}:
\begin{equation}
    S(x^\mu(\tau_1), p_\mu(\tau_1)) = \int_{\tau_0}^{\tau_1} \mathrm{d} \tau \left[ \dot{x}^\mu p_\mu - H(x, p) \right] + \text{const}. 
\end{equation}
Note that the above equation represents an action, with the corresponding Lagrangian related to the Hamiltonian \eqref{eq:H_0} by a Legendre transformation \cite[Ex.~3.6.10]{abraham1978foundations}. The Euler-Lagrange equation is equivalent to the geodesic equation \cite[Th.~3.7.1]{abraham1978foundations} and with Hamilton's equations \eqref{eq:EOM_0_x} and \eqref{eq:EOM_0_p}. 
Once the Hamilton-Jacobi equation is solved, the transport Eq.~\eqref{eq:transp_HJ_0} can also be solved, at least in principle \cite{HJ_transport}. However, our main interest is in the ray equations governed by the Hamiltonian \eqref{eq:H_0}. The corresponding Hamilton's equations \eqref{eq:EOM_0_x} and \eqref{eq:EOM_0_p} describe null geodesics. These equations can easily be rewritten as
\begin{equation}
    \ddot{x}^\mu + \Gamma^\mu_{\alpha \beta} \dot{x}^\alpha \dot{x}^\beta = 0,
\end{equation}
or in the explicitly covariant form:
\begin{equation}
    p^\nu \nabla_\nu p^\mu = \dot{x}^\nu \nabla_\nu \dot{x}^\mu = 0. 
\end{equation}

\subsection{Effective Hamilton-Jacobi system} \label{sec:effective_HJ}

The effective dispersion relation \eqref{eq:eff_disp}, together with the effective transport equation \eqref{eq:eff_transp} introduce $\mathcal{O}(\epsilon^1)$ corrections over the system discussed above:
\begin{align}
    \frac{1}{2} g^{\mu \nu} k_\mu k_\nu - \frac{i\epsilon}{2} k^\mu \left( {a_0}^{*\alpha} \nabla_\mu {a_0}_\alpha - {a_0}_\alpha \nabla_\mu {a_0}^{*\alpha} \right) &= \mathcal{O}(\epsilon^2), \\
    \nabla_\mu \Bigg\{ \mathcal{I} \left[ k^\mu - \frac{i\epsilon}{2} g^{\mu \nu} \left( {a_0}^{*\alpha} \nabla_\nu {a_0}_\alpha - {a_0}_\alpha \nabla_\nu {a_0}^{*\alpha} \right) \right] \Bigg\} &= \mathcal{O}(\epsilon^2).
\end{align}
Using Eq.~\eqref{eq:a0_basis}, the effective dispersion relation becomes
\begin{equation}
    \frac{1}{2} g^{\mu \nu} k_\mu k_\nu - \frac{i\epsilon}{2} k^\mu \left( z^\dagger \partial_\mu z - \partial_\mu z^\dagger z \right) 
    - \epsilon s k^\mu B_\mu = \mathcal{O}(\epsilon^2),
\end{equation}
where $B_\mu = B_\mu(x, k)$ is the Berry connection introduced in Eq.~\eqref{eq:B_connection}, and $s = \pm 1$, depending on the initial polarization. Using Eq.~\eqref{eq:z_integration}, together with the assumption on the initial polarization, we can write:
\begin{equation}
    -\frac{i\epsilon}{2} k^\mu \left( z^\dagger \partial_\mu z - \partial_\mu z^\dagger z \right) = \epsilon s k^\mu \partial_\mu \gamma. 
\end{equation}
Since the value of $s$ is fixed by the initial conditions, the only unknowns are the phase function $S$ and the Berry phase $\gamma$. We can write an effective Hamilton-Jacobi equation for the total phase $\Tilde{S} = S + \epsilon s \gamma$:
\begin{equation} \label{eq:eff_HJ}
\begin{split}
    H\left(x, \nabla \tilde{S} \right) &= \frac{1}{2} g^{\mu \nu} k_\mu k_\nu + \epsilon s k^\mu \partial_\mu \gamma - \epsilon s k^\mu B_\mu + \mathcal{O}(\epsilon^2) \\
    &= \frac{1}{2} g^{\mu \nu} \nabla_\mu \tilde{S} \nabla_\nu \tilde{S} - \epsilon s g^{\mu \nu} B_\mu \nabla_\nu \tilde{S}  + \mathcal{O}(\epsilon^2).
\end{split}
\end{equation}
Note that the phase $\tilde{S}$ represents the overall phase factor, up to order $\mathcal{O}(\epsilon^2)$, of a circularly polarized WKB solution, $\mathcal{A}_\alpha = \mathrm{Re}(\sqrt{\mathcal{I}} m_\alpha e^{i \gamma} e^{i S/\epsilon})$ or $\mathcal{A}_\alpha = \mathrm{Re}(\sqrt{\mathcal{I}} \bar{m}_\alpha e^{-i \gamma} e^{i S/\epsilon})$, depending on the state of circular polarization. As discussed in Ref.~\cite{Bliokh2004}, the Berry phase $\gamma$, which comes as a correction to the overall phase of the WKB solution, is responsible for the spin Hall effect of light.
The corresponding Hamiltonian function on $T^*M$ is
\begin{equation} \label{eq:H_canonical}
    H(x, p) = \frac{1}{2}g^{\mu \nu} p_\mu p_\nu -\epsilon s g^{\mu \nu} p_\mu B_\nu(x, p),
\end{equation}
and we have the following Hamilton's equations:
\begin{align}
    \dot{x}^\mu &= g^{\mu \nu} p_\nu - \epsilon s \left( B^\mu + p^\alpha \vnabla{}^\mu B_\alpha \right), \label{eq:EOM_1_x}\\
    \dot{p}_\mu &=  -\frac{1}{2} \partial_\mu g^{\alpha \beta} p_\alpha p_\beta + \epsilon s p_\alpha \left( \partial_\mu g^{\alpha \beta} B_\beta + g^{\alpha \beta} \partial_\mu B_\beta \right). \label{eq:EOM_1_p}
\end{align}
These equations contain polarization-dependent corrections to the null geodesic Eqs.~\eqref{eq:EOM_0_x} and \eqref{eq:EOM_0_p}, representing the gravitational spin Hall effect of light. For $\epsilon = 0$, one recovers the standard geodesic equation in canonical coordinates. 

We can also write these ray equations in a more compact form
\begin{equation}
    \begin{pmatrix} \dot{x}^\mu \\
    \dot{p}_\mu \end{pmatrix} = \begin{pmatrix} 0 & \delta^\mu_\nu \\
    -\delta^\nu_\mu & 0 \end{pmatrix} \begin{pmatrix} \frac{\partial H}{\partial x^\nu} \\ \frac{\partial H}{\partial p_\nu} \end{pmatrix},
\end{equation}
where the constant matrix on the right-hand side is the inverse of the symplectic $2$-form, or the Poisson tensor \cite{marsden_ratiu}.

\subsubsection{Noncanonical coordinates} \label{sec: noncanonical}

The Hamiltonian \eqref{eq:H_canonical} contains the Berry connection $B_\mu$, which is gauge dependent. The latter means that $B_\mu$ depends on the choice of $m_\alpha$ and $\bar{m}_\alpha$; for example, the transformation $m_\alpha \mapsto m_\alpha e^{i \phi}$ causes the following transformation of the Berry connection:
\begin{equation}
    B_\mu \mapsto B_\mu - \nabla_\mu \phi.
\end{equation}
This kind of gauge dependence was considered by Littlejohn and Flynn in Ref.~\cite{Littlejohn1991}, where they also proposed how to make the Hamiltonian and the equations of motion gauge invariant. The main idea is to introduce noncanonical coordinates such that the Berry connection is removed from the Hamiltonian and the symplectic form acquires the corresponding Berry curvature, which is gauge invariant. This is similar to the description of a charged particle in an electromagnetic field in terms of either the canonical or the kinetic momentum of the particle. The Berry connection and Berry curvature play a similar role as the electromagnetic vector potential and the electromagnetic tensor \cite{Berry_book}.

We start by rewriting the Hamiltonian \eqref{eq:H_canonical} as
\begin{equation}
    H(x, p) = H_0 (x, p) -\epsilon s g^{\mu \nu} p_\mu B_\nu(x, p),
\end{equation}
where $H_0 = \frac{1}{2}g^{\mu \nu} p_\mu p_\nu$. Following Ref.~\cite{Littlejohn1991}, the Berry connection can  be written in the following way, by using the definition of the horizontal derivative:
\begin{equation}
\begin{split}
     p^\mu B_\mu(x, p) &= i p^\mu \bar{m}^{\alpha} \hnabla_\mu {m}_\alpha \\
     &= i p^\mu \bar{m}^{\alpha} \nabla_\mu {m}_\alpha + i p^\mu p_\sigma \Gamma^\sigma_{\mu \rho} \bar{m}^{\alpha} \vnabla{}^\rho {m}_\alpha  \\
     &= i \frac{\partial H_0}{ \partial p_\mu} \bar{m}^{\alpha} \nabla_\mu {m}_\alpha - i \frac{\partial H_0}{ \partial x^\mu} \bar{m}^{\alpha} \vnabla{}^\rho {m}_\alpha.
\end{split}
\end{equation}

The Berry connection can be eliminated formally from the Hamiltonian \eqref{eq:H_canonical} by considering the following substitution on $T^*M$: 
\begin{align} \label{eq:coord}
    X^\mu &= x^\mu + i \epsilon s \bar{m}^{\alpha} \vnabla{}^\mu {m}_\alpha, \\
    P_\mu &= p_\mu - i \epsilon s \bar{m}^{\alpha} \nabla_\mu {m}_\alpha. \label{eq:coord1}
\end{align}
It is possible to obtain this substitution as the linearization of a change of coordinates. For more details, see Appendix~\ref{app:coord}.

Since the symplectic form transforms nontrivially under this substitution, $(X, P)$ are noncanonical coordinates. The Hamiltonian \eqref{eq:H_canonical} is a scalar, so we obtain
\begin{equation}
\begin{split}
    &H'(X, P) = H(x, p) \\
    &= H \left( X^\mu - i \epsilon s \bar{m}^{\alpha} \vnabla{}^\mu {m}_\alpha, P_\mu + i \epsilon s \bar{m}^{\alpha} \nabla_\mu {m}_\alpha \right) \\
    &= H(X, P) - i \epsilon s \frac{\partial H_0}{ \partial x^\mu} \bar{m}^{\alpha} \vnabla{}^\mu {m}_\alpha + i \epsilon s \frac{\partial H_0}{ \partial p_\mu} \bar{m}^{\alpha} \nabla_\mu {m}_\alpha \\
    &= H_0(X, P).
\end{split}
\end{equation}
In the new coordinate system $(X, P)$, we obtain the following Hamiltonian:
\begin{equation}\label{eq:coordinvham}
    H'(X, P) = \frac{1}{2}g^{\mu \nu}(X) P_\mu P_\nu.
\end{equation}
The corresponding Hamilton's equations can be written in a matrix form as
\begin{equation}
    \begin{pmatrix} \dot{X}^\mu \\
    \dot{P}_\mu \end{pmatrix} = T' \begin{pmatrix} \frac{\partial H'}{\partial X^\nu} \\ \frac{\partial H'}{\partial P_\nu} \end{pmatrix},
\end{equation}
where $T'$ is the Poisson tensor in the new variables. Following Marsden and Ratiu \cite[p. 343]{marsden_ratiu}, we obtain
\begin{equation}
    T' = \begin{pmatrix} \epsilon s  \left( F_{p p} \right)^{\nu \mu} & \delta^\mu_\nu + \epsilon s \left(F_{x p} \right)\indices{_\nu^\mu} \\
    -\delta^\nu_\mu - \epsilon s \left(F_{x p} \right)\indices{^\nu_\mu} & - \epsilon s \left(F_{x x} \right)_{\nu \mu} \end{pmatrix},
\end{equation}
where we have the following Berry curvature terms:
\begin{equation} \label{eq:Berry_curvature}
\begin{split}
    &\begin{split} \left({F_{p p}}\right)^{\nu \mu} = i \Big( &\vnabla{}^\mu \bar{m}^\alpha  \vnabla{}^\nu m_\alpha - \vnabla{}^\nu \bar{m}^\alpha \vnabla{}^\mu m_\alpha \\ &+ \bar{m}^\alpha \vnabla{}^{[\mu} \vnabla{}^{\nu]} m_\alpha - m_\alpha \vnabla{}^{[\mu} \vnabla{}^{\nu]} \bar{m}^\alpha \Big), \end{split} \\
    &\begin{split} \left({F_{xx}}\right)_{\nu \mu} = i \Big( &\nabla_\mu \bar{m}^\alpha \nabla_\nu m_\alpha - \nabla_\nu \bar{m}^\alpha \nabla_\mu m_\alpha \\ &+ \bar{m}^\alpha \nabla_{[\mu} \nabla_{\nu]} m_\alpha - m_\alpha \nabla_{[\mu} \nabla_{\nu]} \bar{m}^\alpha \Big), \end{split} \\
    &\begin{split}\left({F_{p x}}\right)\indices{_\nu^\mu} &= -\left({F_{x p}}\right)\indices{^\mu_\nu} \\
    &= i \left( \vnabla{}^\mu \bar{m}^\alpha \nabla_\nu m_\alpha - \nabla_\nu \bar{m}^\alpha \vnabla{}^\mu m_\alpha \right).\end{split}
\end{split}
\end{equation}
The Poisson tensor in noncanonical coordinates $T'$ automatically satisfies the Jacobi identity, since it is a covariant quantity obtained from the Poisson tensor in canonical coordinates $T$ through a change of variables on the cotangent bundle. 

Simplified expressions for the Berry curvature terms can be found in Appendix~\ref{app:Berry_curvature}. Now we can write Hamilton's equations in the new variables:
\begin{align}
    \begin{split}\dot{X}^\mu \, &= \, P^\mu + \epsilon s P^\nu \left( F_{p x} \right)\indices{_\nu^\mu} \\
    &\qquad\qquad+ \epsilon s \Gamma^\alpha_{\beta \nu} P_\alpha P^\beta \left( F_{p p}\right)^{\nu \mu} ,\end{split} \label{eq:Xdot}  \\
    \begin{split}\dot{P}_\mu \, &= \, \Gamma^\alpha_{\beta \mu} P_\alpha P^\beta - \epsilon s P^\nu \left( F_{x x} \right)_{\nu \mu} \\
    &\qquad\qquad- \epsilon s \Gamma^\alpha_{\beta \nu} P_\alpha P^\beta \left( F_{x p} \right)\indices{^\nu_\mu} .\end{split} \label{eq:Pdot}
\end{align}
The last term on the right-hand side of Eq.~\eqref{eq:Xdot} is the covariant analogue of the spin Hall effect correction obtained in optics, $\left( \dot{\mathbf{p}} \times \mathbf{p}\right) / |\mathbf{p}|^3$, due to the Berry curvature in momentum space \cite{SOI_review,Ruiz2015}. This term is also the source of the gravitational spin Hall effect in the work of Gosselin \textit{et~al.} \cite{SHE_QM1}. In Eq.~\eqref{eq:Pdot}, the second term on the right-hand side contains the Riemann tensor and resembles the curvature term obtained in the Mathisson-Papapetrou-Dixon equations \cite{dixon2015new}. 

Given a null covector $P_\mu$, the class of Lorentz transformations leaving $P_\mu$ invariant define the little group, which is isomorphic to $\textrm{SE}(2)$, the symmetry group of the two-dimensional Euclidean plane \cite{Stone2015}. In terms of a null tetrad $\{ P, n, m, \bar{m} \}$, the action of the little group can be split into the following types of transformations \cite[p. 53]{Chandra}:
\begin{equation}
\begin{split}
    \text{Type 1:}& \qquad P \mapsto P, \quad n \mapsto n,\\
    &\qquad m \mapsto m e^{i \phi}, \quad m \mapsto \bar{m} e^{-i \phi}, \\
    \text{Type 2:}& \qquad P \mapsto P, \quad n \mapsto n + \bar{a}m + a \bar{m} + a \bar{a} P,\\
    &\qquad m \mapsto m + a P, \quad m \mapsto \bar{m} + \bar{a} P,
\end{split}
\end{equation}
where $\phi$ is a real scalar function and $a$ is a complex scalar function. The transformations of Type 1 are the spin rotations mentioned in Sec.~\ref{sec:polarization_basis}, while the transformations of Type 2 can be considered as a change of observer $t_\mu$, based on Eq.~\eqref{eq:t2n}. It can easily be checked that the Berry curvature terms in Eq.~\eqref{eq:Berry_curvature} are invariant under Type 1 transformations. However, the Berry curvature terms are not invariant under Type 2 transformations. As a consequence, the ray equations \eqref{eq:Xdot} and \eqref{eq:Pdot} depend on the choice of observer. It is shown in the following section how this observer dependence is related to the problem of localizing massless spinning particles \cite{Relativistic_Hall,Stone2015}.

\section{Examples} \label{sec:numerics}

In this section, we apply the modified ray equations describing the gravitational spin Hall effect of light to two concrete examples. The first example, concerning the relativistic Hall effect and Wigner translations, is treated analytically, while the second example, describing the propagation of polarized light rays close to a Schwarzschild black hole, is treated numerically.

When working with the modified ray equations, in either the canonical form given in Eqs.~\eqref{eq:EOM_1_x} and \eqref{eq:EOM_1_p} or the noncanonical form given in Eqs.~\eqref{eq:Xdot} and \eqref{eq:Pdot}, one needs to specify the background metric $g_{\mu \nu}$, and the choice of polarization vectors $m^\alpha$ and $\bar{m}^\alpha$. The polarization vectors are needed in order to compute the Berry connection and the Berry curvature. A particular choice of polarization vectors can easily be constructed by introducing an orthonormal tetrad $(e_a)^\mu$, with $(e_0)^\mu = t^\mu$ representing our choice of family of timelike observers. Adapting the polarization vectors used in optics \cite{Ruiz2015}, we can write $p^\mu = P^a (e_a)^\mu$, $v^\mu = V^a(e_a)^\mu$, and $w^\mu = W^a (e_a)^\mu$, where the components of these vectors are given by
\begin{equation} \label{eq:polarization_vectors}
\begin{split}
    P^a &= \begin{pmatrix} 
    			P^0 \\ P^1 \\ P^2 \\ P^3
   		 \end{pmatrix}, \qquad 
    V^a = \frac{1}{P_p} \begin{pmatrix} 
    			0 \\ -P^2 \\ P^1 \\ 0
    		\end{pmatrix},\\
    W^a &= \frac{1}{P_p P_s} \begin{pmatrix} 
    			0 \\ P^1 P^3 \\ P^2 P^3 \\ -(P_p)^2
    \end{pmatrix},
\end{split}
\end{equation}
where
\begin{equation}
\begin{split}
    P_p &= \sqrt{\left(P^1 \right)^2 + \left(P^2 \right)^2},\\
    P_s &= \sqrt{\left(P^1 \right)^2 + \left(P^2 \right)^2 + \left(P^3 \right)^2}.
\end{split}
\end{equation}
The vectors $v^\mu$ and $w^\mu$ are real unit spacelike vectors that represent a linear polarization basis satisfying Eq.~\eqref{eq:tetrad_prop}. They are related to the circular polarization vectors $m^\alpha$ and $\bar{m}^\alpha$ by Eq.~\eqref{eq:circ-to-linear}. Using this particular choice of polarization vectors, the Berry connection and the Berry curvature terms can be computed, and the modified ray equations can be integrated, either analytically or numerically.

\subsection{Relativistic Hall effect and Wigner translations} \label{sec:Wigner}

The relativistic Hall effect \cite{Relativistic_Hall} is a special relativistic effect that occurs when Lorentz transformations are applied to objects carrying angular momentum. In particular, consider a localized wave packet carrying intrinsic angular momentum and propagating in the $z$ direction in Minkowski spacetime. If a Lorentz boost is applied in the $x$ direction, then the location of the Lorentz-transformed energy density centroid is shifted in the $y$ direction, depending on the orientation of the angular momentum. This shift corresponds to the Wigner translation \cite{Stone2015,DUVAL2015322,BOLONEKLASON2017117}.

The following example shows that an effect analogous to the Wigner translation discussed in Ref.~\cite{Stone2015} appears in the effective ray equations \eqref{eq:Xdot} and \eqref{eq:Pdot}. We consider the Minkowski spacetime in Cartesian coordinates $(t, x, y, z)$, with
\begin{equation} \label{eq:std_tetrad}
    \mathrm{d}s^2 = -\mathrm{d}t^2 + \mathrm{d}x^2 + \mathrm{d}y^2 + \mathrm{d}z^2,
\end{equation}
and we want to compare the effective rays obtained from Eqs.~\eqref{eq:Xdot} and \eqref{eq:Pdot} with two different choices of observer. In the first case, we consider the standard orthonormal tetrad
\begin{equation}
    e_0 = \partial_t, \qquad e_1 = \partial_x, \qquad e_2 = \partial_y, \qquad e_3 = \partial_z,
\end{equation}
where $(e_0)^\mu$ is our first choice of observer. With this orthonormal tetrad, the polarization vectors are defined as in Eq.~\eqref{eq:polarization_vectors}, and the Berry curvature terms can be computed. The ray equations reduce to the geodesic equations
\begin{equation}
    \dot{X}^\mu = P^\mu, \qquad  \dot{P}_\mu = 0.
\end{equation}
In order to describe light rays traveling in the $z$ direction, we impose initial conditions $X^\mu (0) = (0,0,0,0)$ and $P_\mu (0) = (-1,0,0,1)$, and we obtain
\begin{equation}
\begin{split}
    X^\mu(\tau) &= \left( \tau, 0, 0, \tau \right), \\
    P_\mu(\tau) &= \left( -1, 0, 0, 1 \right).
\end{split}
\end{equation}
As a second case, we apply a time-dependent boost in the $x$ direction to the standard orthonormal tetrad in Eq.~\eqref{eq:std_tetrad}. We obtain
\begin{equation} \label{eq:booster_tetrad}
\begin{alignedat}{2}
    e'_0 &= \phantom{+}\cosh t \, \partial_t - \sinh t \, \partial_x, \qquad  && e'_2 = \partial_y \\
    e'_1 &= -\sinh t \, \partial_t + \cosh t \, \partial_x,  \qquad && e'_3 = \partial_z,
\end{alignedat}
\end{equation}
where $(e'_0)^\mu$ is our second choice of observer. Note that $(e'_0)^\mu$ represents a family of observers boosted in the $x$ direction, with the rapidity of the boost represented by the time coordinate $t$. The polarization vectors are chosen as in Eq.~\eqref{eq:polarization_vectors}, but this time with respect to the orthonormal tetrad in Eq.~\eqref{eq:booster_tetrad}. The Berry curvature terms in Eqs.~\eqref{eq:Xdot} and \eqref{eq:Pdot} can be explicitly computed, and we obtain
\begin{figure*}[t!]
\subfigure{\includegraphics[width=.53\textwidth]{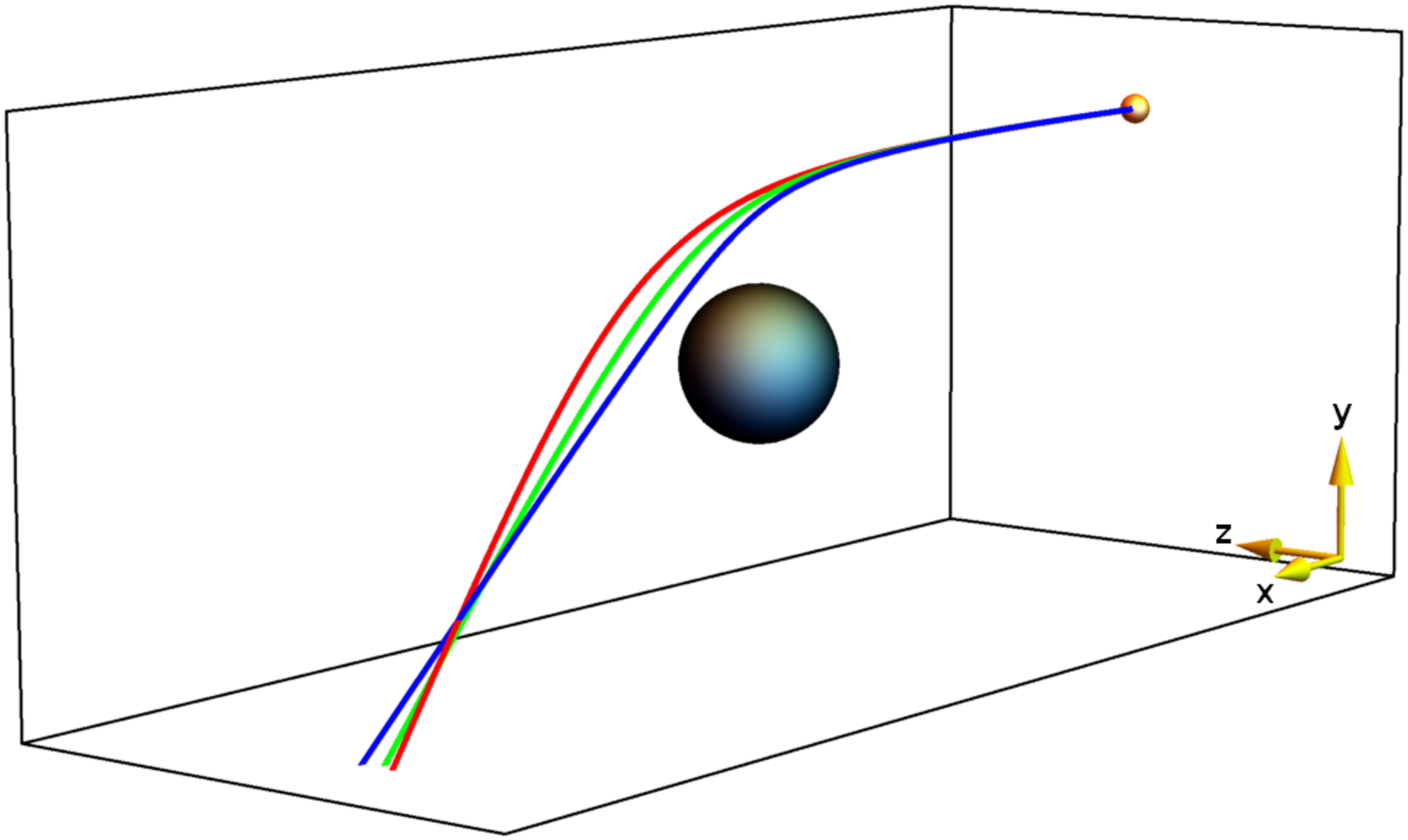}} \hspace{1.5cm}
\subfigure{\includegraphics[width=.30\textwidth]{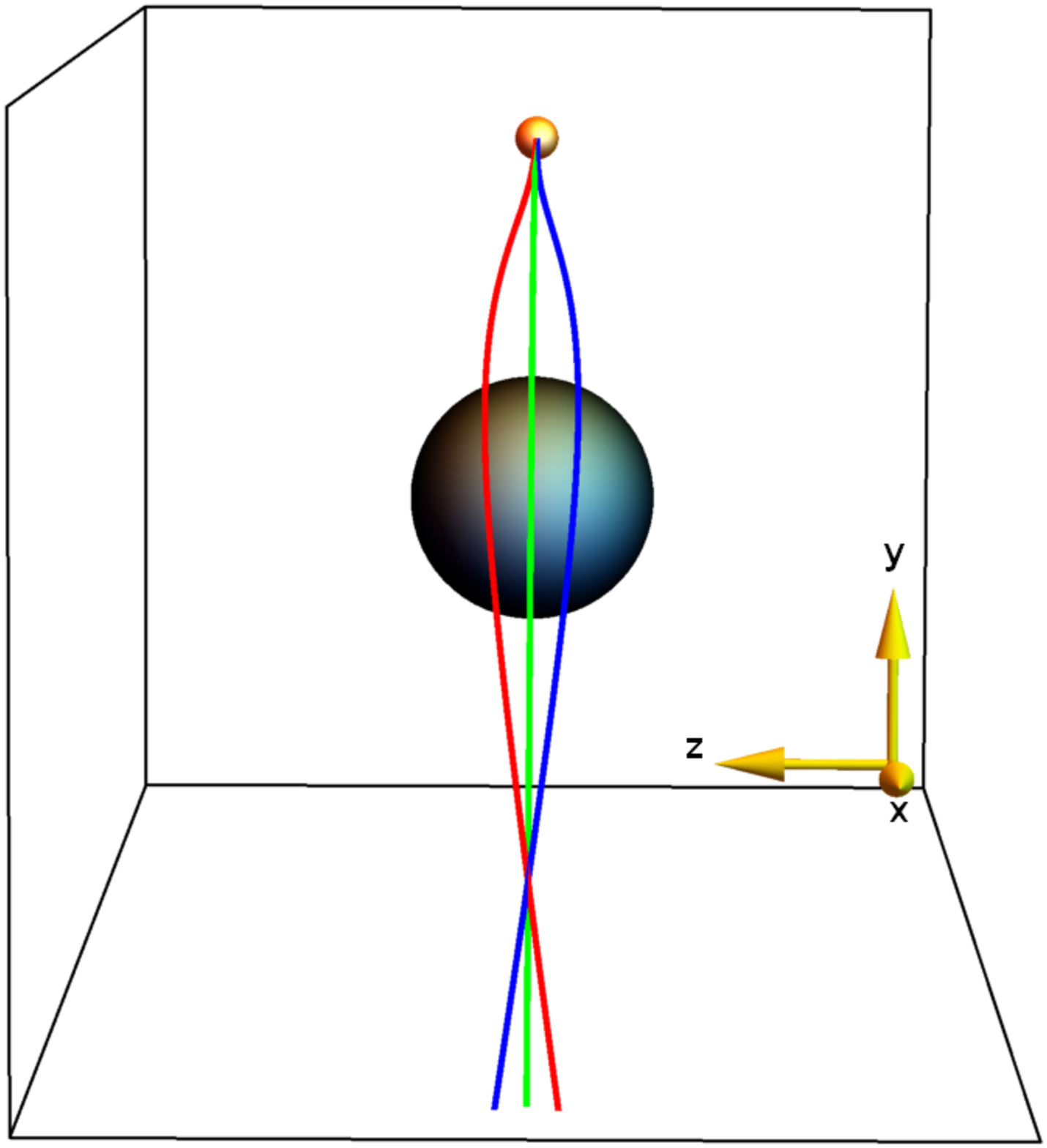}} 
\caption{Results of numerical simulations illustrating the gravitational spin Hall effect of light around a Schwarzschild black hole. The effect is exaggerated for visualization purposes. The two figures present the same rays from different viewing angles. The central sphere represents the Schwarzschild black hole, and the small orange sphere represents a source of light. The blue and the red trajectories correspond to rays of opposite circular polarization, $s= \pm 1$, while the green trajectory represents a null geodesic. We take $r_s = 1$, and we  start with the initial position $x^i(0) = (-50 r_s, 15 r_s, 0)$, and initial normalized momentum $p_i = (1, 0, 0)$. The wavelength $\lambda$ is set to a sufficiently large value to make the effect visible on this plot.}
\label{fig:GSHE}
\end{figure*}
\begin{align}
    \dot{X}^\mu &= P^\mu + \epsilon s P^\nu \left( F_{p x} \right)\indices{_\nu^\mu}, \\
    \dot{P}_\mu &= 0,
\end{align}
where
\begin{equation}
     P^\nu \left( F_{p x} \right)\indices{_\nu^\mu} = \frac{P_t}{ \left[ (e'_0)^\mu P_\mu \right]^2 } \begin{pmatrix} 
    			0 \\ 0 \\ P_z \\ -P_y
   		 \end{pmatrix}.
\end{equation}
We impose the same initial conditions as in the previous case: $X^\mu (0) = (0,0,0,0)$ and $P_\mu (0) = (-1,0,0,1)$. Since the frequency is defined as $\omega = - (e'_0)^\mu P_\mu/ \epsilon$, the small parameter $\epsilon$ can be identified with the wavelength of the initial light ray, as measured by the observer $(e'_0)^\mu$ at the spacetime point $x^\mu = X^\mu(0)$. Then, the ray equations can be analytically integrated, and we obtain
\begin{equation}
\begin{split}
    X^\mu(\tau) &= \left( \tau, 0, -s \epsilon \tanh \tau, \tau \right), \\
    P_\mu(\tau) &= \left( -1, 0, 0, 1 \right).
\end{split}
\end{equation}
Thus, given a circularly polarized light ray traveling in the $z$ direction and two families of observers $(e_0)^\mu$ and $(e'_0)^\mu$, which are related by boosts in the $x$ direction, we obtained the polarization-dependent Wigner translation in the $y$ direction, $\Delta y = s \epsilon \tanh \tau$, in agreement with \cite[Eq.~28]{Stone2015}. Note that the Wigner translation is always smaller than one wavelength.

Recovering the results of Ref.~\cite{Stone2015} suggests that a worldline $X^\mu(\tau)$ representing a solution of Eqs.~\eqref{eq:Xdot} and \eqref{eq:Pdot} could be interpreted as the location of the energy density centroid of a localized wave packed with definite circular polarization, as measured by the chosen family of observers.

\subsection{The gravitational spin Hall effect on a Schwarzschild background}

To illustrate how the polarization-dependent correction terms modify the ray trajectories on a Schwarzschild background, let us provide some numerical examples. For convenience, we perform the numerical computations using canonical coordinates $(x, p)$ and treat $x^0$ as a parameter along the rays. Hence, Eqs.~\eqref{eq:EOM_1_x} and \eqref{eq:EOM_1_p} become
\begin{align}
    \dot{x}^0&= 1, \label{eq:EOM_tparam1} \\
    \dot{x}^i &= \dfrac{g^{i \nu} p_\nu - \epsilon s \left( B^i + p^\alpha \vnabla{}^i B_\alpha \right)}{g^{0 \nu} p_\nu - \epsilon s \left( B^0 + p^\alpha \vnabla{}^0 B_\alpha \right)}, \label{eq:EOM_tparam2}\\
    \dot{p}_i &=  \dfrac{-\frac{1}{2} \partial_i g^{\alpha \beta} p_\alpha p_\beta + \epsilon s p_\alpha \left( \partial_i g^{\alpha \beta} B_\beta + g^{\alpha \beta} \partial_i B_\beta \right)}{g^{0 \nu} p_\nu - \epsilon s \left( B^0 + p^\alpha \vnabla{}^0 B_\alpha \right)}, \label{eq:EOM_tparam3}
\end{align}
and $p_0$ is calculated from
\begin{equation}
\dfrac{1}{2} g^{\mu \nu}p_\mu p_\nu - \epsilon s g^{\mu \nu} p_\mu B_\nu(x, p) = 0.
\end{equation}
This equation can be solved explicitly, using the fact that the velocity $\dot{x}^\alpha$ is future oriented:
\begin{equation}
\begin{split}
&p_0 = \frac{1}{g^{00}} \Bigg[-\left(g^{0i}p_i - \epsilon s g^{0\mu}B_\mu\right) \\
+& \sqrt{\left(g^{0i}p_i - \epsilon s g^{0\mu}B_\mu\right)^2 - g^{00}\left( g^{ij}p_ip_j -2\epsilon s p_ig^{i\mu}B_\mu\right)}\Bigg].
\end{split}
\end{equation}
Note that in general $B_\mu$ depends on $p_0$. However, since this is an $\mathcal{O}(\epsilon^1)$ term, we can replace the $\mathcal{O}(\epsilon^0)$ expression for $p_0$ in $B_\mu$.

In order to compare with the results of Gosselin \textit{et~al.} \cite{SHE_QM1}, we consider a Schwarzschild spacetime in Cartesian isotropic coordinates $(t, x, y, z)$:
\begin{equation}
    \mathrm{d}s^2 = - \left( \frac{1 - \frac{r_s}{4R}}{1 + \frac{r_s}{4R}} \right)^2 \mathrm{d}t^2 + \left( {1 + \frac{r_s}{4R}} \right)^4 (\mathrm{d}x^2 + \mathrm{d}y^2 + \mathrm{d}z^2),
\end{equation}
where $r_s = 2 G M/c^2$ is the Schwarzschild radius and $R = \sqrt{x^2+y^2+z^2}$. We also define the following orthonormal tetrad:
\begin{equation} \label{eq:orth_tetrad}
\begin{alignedat}{2}
    e_0 &= \frac{1 + \frac{r_s}{4R}}{1 - \frac{r_s}{4R}} \, \partial_t,   &&e_1 = \left( {1 + \frac{r_s}{4R}} \right)^{-2} \, \partial_x, \\
    e_2 &= \left( {1 + \frac{r_s}{4R}} \right)^{-2} \, \partial_y,  \quad &&e_3 = \left( {1 + \frac{r_s}{4R}} \right)^{-2} \, \partial_z,
\end{alignedat}
\end{equation}
where $t^\mu = (e_0)^\mu$ is our choice of observer.
\begin{figure*}[t!]
    \includegraphics[width=.70\textwidth]{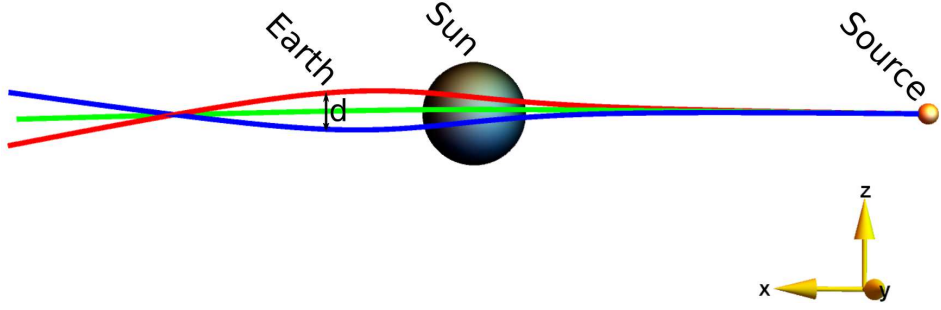}
\caption{Results of numerical simulations illustrating the gravitational spin Hall effect of light around the Sun. The effect is exaggerated for visualization purposes. The separation distance $d$ is observed from the Earth. The blue and the red trajectories correspond to rays of opposite circular polarization, $s= \pm 1$, while the green trajectory represents a null geodesic. We take $r_s = 3 ~\mathrm{km}$, and we start with the initial position $x^i(0) = (-10^7 r_s, 3 \times 10^5 r_s, 0)$, and initial normalized momentum $p_i = (1, 0, 0)$.}
\label{fig:GSHE2}
\end{figure*}

The Berry connection $B_\mu$ can be explicitly computed by introducing a particular choice of polarization vectors, using Eq.~\eqref{eq:polarization_vectors} and the orthonormal tetrad \eqref{eq:orth_tetrad}. We now have all the elements required for the numerical integration of Eqs.~\eqref{eq:EOM_tparam1}-\eqref{eq:EOM_tparam3}. For this purpose, we used the \textsc{NDSolve} function of \textit{Mathematica} \cite{Mathematica}. For these examples, we used the default settings for integration method, precision and accuracy. 

After obtaining a numerical solution $(x(t),p(t))$ to Eqs.~\eqref{eq:EOM_tparam1}-\eqref{eq:EOM_tparam3}, in order to ensure the gauge invariance of our results, we have to evaluate the gauge-invariant noncanonical quantities $(X(t), P(t))$, as given in Eqs.~\eqref{eq:coord} and \eqref{eq:coord1}. These are the quantities used to represent the trajectories in Figs.~\ref{fig:GSHE} and \ref{fig:GSHE2}. A comparative discussion between the use of canonical and noncanonical ray equations in optics, together with numerical examples, can be found in Ref.~\cite{Ruiz2015}.

As the first step, we numerically compare our ray Eqs.~\eqref{eq:EOM_tparam1}-\eqref{eq:EOM_tparam3} with those predicted by Gosselin \textit{et~al.} \cite{SHE_QM1}. This is done by numerically integrating Eqs.~\eqref{eq:EOM_tparam1}-\eqref{eq:EOM_tparam3}, as well as Eq.~(23) from Ref.~\cite{SHE_QM1}. Up to numerical errors, we obtain the same ray trajectories with both sets of equations. However, while the equations obtained by Gosselin \textit{et~al.} only apply to static spacetimes, Eqs.~\eqref{eq:EOM_tparam1}-\eqref{eq:EOM_tparam3} do not have this limitation.

The results of our numerical simulations are shown in Fig.~\ref{fig:GSHE}, which illustrates the general behavior of the gravitational spin Hall effect of light around a Schwarzschild black hole. [The actual effect is small, so the figure is obtained by numerical integration of Eqs.~\eqref{eq:EOM_tparam1}-\eqref{eq:EOM_tparam3} for unrealistic parameters.] Here, we consider rays of opposite circular polarization ($s = \pm 1$) passing close to a Schwarzschild black hole, together with a reference null geodesic ($s = 0$). Except for the value of $s$, we are considering the same initial conditions, $\left( x^i(0), p_i(0) \right)$, for these rays. Unlike the null geodesic, for which the motion is planar, the circularly polarized rays are not confined to a plane. 

As another example, we used initial conditions $\left( x^i(0), p_i(0) \right)$ such that the rays are initialized as radially ingoing or outgoing. In this case (not illustrated, since it is trivial), the gravitational spin Hall effect vanishes, and the circularly polarized rays coincide with the radial null geodesic.  

Using these numerical methods, we can also estimate the magnitude of the gravitational spin Hall effect. As a particular example, we consider a similar situation to the one presented in Fig.~\ref{fig:GSHE}, where the black hole is replaced with the Sun. More precisely, we model this situation by considering a Schwarzschild black hole with $r_s \approx 3~\mathrm{km}$. We consider the deflection of circularly polarized rays coming from a light source far away, passing close to the surface of the Sun, and then observed on the Earth. This situation is illustrated in Fig.~\ref{fig:GSHE2}. The numerical results are based on the initial data presented in the caption of Fig.~\ref{fig:GSHE2}. When reaching the Earth, the separation distance between the rays of opposite circular polarization depends on the wavelength. For example, taking wavelengths of the order $\lambda \approx 10^{-9}~\mathrm{m}$ results in a separation distance of the order $d \approx 10^{-15}~\mathrm{m}$, while for wavelengths of the order of $\lambda \approx 1~\mathrm{m}$ we obtain a separation distance of the order $d \approx 10^{-6}~\mathrm{m}$. Although the ray separation is small (about six orders of magnitude smaller than the wavelength), what really matters is that the rays are scattered by a finite angle. Therefore, the ray separation grows linearly with distance after the reintersection point. This means that the effect should be robustly observable if one measures it sufficiently far from the Sun. Furthermore, massive compact astronomical objects, such as black holes or neutron stars, are expected to produce a larger gravitational spin Hall effect. 

As a consistency check, we also performed the numerical computations using different coordinates, such as the standard Schwarzschild spherical coordinates and Gullstrand–Painlev\'e coordinates. The results are independent of the choice of coordinates. However, the polarized rays are not invariant under a change of observer. This is due to an effect analogous to the Wigner translations discussed in Sec.~\ref{sec:Wigner}. For example, instead of the static observer introduced in Eq.~\eqref{eq:orth_tetrad}, one could consider a free-falling observer. In this case, the ray trajectories presented in Figs. \ref{fig:GSHE} and \ref{fig:GSHE2} are slightly modified, due to the Wigner translations, and preliminary investigations indicate that these modifications are smaller than one wavelength, as in the case discussed in Sec.~\ref{sec:Wigner}. It is not clear how to separate the purely gravitational effect from the observer-dependent Wigner translations. However, this is not a problem. The Wigner translation represents the observer-dependent ambiguity in defining the location of the ray on a single-wavelength scale and remains bounded. In contrast, the purely gravitational effect can affect the angle of light scattering off a gravitating object and thus the ray displacement associated with this effect accumulates linearly with the distance. This means that the latter effect dominates at large distances. A more detailed analysis of the modified ray equations, at both the analytical and the numerical levels, will be carried out in future work.

\section{Conclusions} \label{sec:conclusions}

In summary, we have presented a first comprehensive theory of the gravitational spin Hall effect that occurs due to the coupling of the polarization with the translational dynamics of the light rays. The ray dynamics is governed by the corrected Hamiltonian
\begin{equation} 
     H(x, p) = \frac{1}{2}g^{\mu \nu} p_\mu p_\nu -\epsilon s g^{\mu \nu} p_\mu B_\nu(x, p).
\end{equation}
Here, the first term represents the geometrical optics Hamiltonian, and the second term represents a correction of $\mathcal{O}(\epsilon^1)$ that is due to the Berry connection, which is given by
\begin{equation}
\begin{split}
 B_\mu(x, p) &= i \bar{m}^{\alpha}\left(\frac{\partial}{\partial x^\mu} m_\alpha - \Gamma^\sigma_{\alpha \mu} m_\sigma + \Gamma^\sigma_{\mu \rho} p_\sigma \frac{\partial}{\partial p_\rho} m_\alpha \right)\\
 &= i \bar{m}^{\alpha} \hnabla_\mu {m}_\alpha.
\end{split}
\end{equation}
Assuming the noncanonical coordinates \eqref{eq:coord}, the corresponding ray equations are
 \begin{align}
    \begin{split}
     \dot{X}^\mu \, &= \, P^\mu + \epsilon s P^\nu \left( F_{p x} \right)\indices{_\nu^\mu} \\
      &\qquad\qquad+ \frac{2i\epsilon s}{{(t^\alpha P_\alpha)^2}} \Gamma^\alpha_{\beta \nu} P_\alpha P^\beta m^{[\nu} \bar{m}^{\mu]}, \end{split} \label{eq:Xdot1} \\
     \begin{split} 
     \dot{P}_\mu \, &= \, \Gamma^\alpha_{\beta \mu} P_\alpha P^\beta + \epsilon s P^\nu \left[  i R_{\alpha \beta \mu \nu} m^\alpha \bar{m}^\beta + (\tilde{F}_{x x}){}_{\nu \mu} \right] \\
     &\qquad\qquad+ \epsilon s \Gamma^\alpha_{\beta \nu} P_\alpha P^\beta \left( F_{p x} \right)\indices{_\mu^\nu}, \end{split} \label{eq:Pdot1}
 \end{align}
where the terms $F_{px}$ and $\tilde{F}_{x x}$ and the timelike vector $t^\alpha$ are given in Appendix~\ref{app:Berry_curvature}. The last term on the right-hand side of Eq.~\eqref{eq:Xdot1} is the covariant analogue of the spin Hall correction term usually encountered in optics \cite{Bliokh2008,SHE_QM1}, while the Riemann curvature term in Eq.~\eqref{eq:Pdot1} is reminiscent of a similar term appearing in the Mathisson-Papapetrou-Dixon equations \cite{dixon2015new}. In Minkowski spacetime, the  $F_{px}$ term is responsible for the relativistic Hall effect \cite{Relativistic_Hall} and Wigner translations \cite{Stone2015}.

The resulting deviation of the ray trajectories from those predicted by geometrical optics is weak but not unobservable. First of all, even small angular deviations are observable at large enough distances. Second, as shown in Ref.~\cite{SHEL_experiment}, weak quantum measurement techniques can be used to detect the spin Hall effect of light, even when the spatial separation between the left-polarized and the right-polarized beams of light is smaller than the wavelength.

Potentially, this work can be naturally extended in two directions. First, the corrected ray equations are yet to be studied more thoroughly, both analytically and numerically. Rigorous numerical investigations are needed to obtain a precise prediction of the effect, in particular for Kerr black holes. Second, Maxwell's equations are a proxy to linearized gravity. It is expected that a similar approach can be carried out to obtain an effective pointwise description of a gravitational wave packet, extending the results of Ref.~\cite{SHE_GW}. 

As discussed in Ref.~\cite{SOI_review}, the spin Hall effect of light is directly related to the conservation of total angular momentum. For the discussion presented so far, the considered rays carry extrinsic orbital angular momentum, associated with the ray trajectory, and intrinsic spin angular momentum, associated with the polarization. However, it is well-known that light can also carry intrinsic orbital angular momentum \cite{Allen92,AM_Light,GW_IOAM2} (see also Ref.~\cite{2018arXiv181203292A} and references therein). In principle, the magnitude of the spin Hall effect can be increased by considering optical beams carrying intrinsic orbital angular momentum \cite{Bliokh2006}. The method and ansatz that we have adopted are insufficient to describe this effect. A more realistic and more precise approach involving wave packets, such as Laguerre-Gaussian beams, should be considered. It may be possible to do so using the machinery developed in Ref.~\cite{Dodin2018}.

A formulation of the special-relativistic dynamics of massless spinning particles and wave packets beyond the geometrical optics limit has been previously reported by Duval and collaborators ( cf.~Ref.~\cite{2015PhLB..742..322D} for the spin-$1/2$ case; see also Ref.~\cite{2016IJMPB..3050249S}). This analysis relates the modified dynamics to the approach of Souriau \cite{MR1461545}, making use of so-called spin enslaving. This has been extended to general helicity by Andrzejewski \textit{et~al.} \cite{2015PhLB..746..417A}. We expect that the Hamiltonian formulation presented here corresponds to a general relativistic version of the models considered in the mentioned papers. This will be considered in a future work.

\subsection*{Acknowledgements}

We are grateful to Pedro Cunha for helpful discussions. A significant part of this work was done while one of the authors, L.A., was in residence at Institut Mittag-Leffler in Djursholm, Sweden during the fall of 2019, supported by the Swedish Research Council under Grant no. 2016-06596. I.Y.D. acknowledges support from the U.S. National Science Foundation under Grant No. PHY 1903130. M.A.O. is supported by the International Max Planck Research School for Mathematical and Physical Aspects of Gravitation, Cosmology and Quantum Field Theory. C.F.P. was partially supported by the Australian Research Council Grant no. DP170100630 and partially funded by the SNSF Grant no. P2SKP2 178198.

Sandia National Laboratories is a multimission laboratory managed and operated by the National Technology \& Engineering Solutions of Sandia, LLC, a wholly owned subsidiary of Honeywell International Inc., for the U.S. Department of Energy (DOE) National Nuclear Security Administration under Contract No. DE-NA0003525. This paper describes the objective technical results and analysis. Any subjective views or opinions that might be expressed in the paper do not necessarily represent the views of the U.S. DOE or the United States Government.

\appendix
\renewcommand{\theequation}{\thesection\arabic{equation}}

\section{Horizontal and vertical derivatives on $\boldsymbol{T^*M}$} \label{app:derivative_TM}

Let $(x^\mu, p_\mu)$ be canonical coordinates on $T^*M$. Considering fields defined on $T^*M$, such as $u_\alpha(x, p)$ and $v^\alpha(x, p)$, the horizontal and vertical derivatives are defined as follows \cite[Sec.~3.5]{sharafutdinov2012integral}:
\begin{subequations} 
\begin{align}
\vnabla{}^\mu u_\alpha ={}& \frac{\partial}{\partial p_\mu} u_\alpha, \\
\hnabla_\mu u_\alpha ={}&  
\frac{\partial}{\partial x^\mu} u_\alpha - \Gamma^\sigma_{\alpha \mu} u_\sigma + \Gamma^\sigma_{\mu \rho} p_\sigma \frac{\partial}{\partial p_\rho} u_\alpha,
\end{align} 
\end{subequations} 
\begin{subequations} 
\begin{align}
\vnabla{}^\mu v^\alpha ={}& \frac{\partial}{\partial p_\mu} v^\alpha, \\
\hnabla_\mu v^\alpha ={}&  
\frac{\partial}{\partial x^a} v^\alpha + \Gamma^\alpha_{\sigma \mu} v^\sigma + \Gamma^\sigma_{\mu \rho} p_\sigma \frac{\partial}{\partial p_\rho} v^\alpha. 
\end{align} 
\end{subequations} 
The extension to general tensor fields on $T^*M$ is straightforward. Note that in contrast to Ref.~\cite[Sec.~3.5]{sharafutdinov2012integral}, we have the opposite sign for the last term in the definition of the horizontal derivative. This is because our fields, $u_\alpha(x, p)$ and $v^\alpha(x, p)$, are defined on $T^*M$, and not on $TM$, as is the case in the reference mentioned before. We can make use of the following properties:
\begin{equation} \label{eq:hor_vert_prop}
\begin{split}
    [\hnabla{}_\mu, \vnabla{}^\nu] =0, \quad [\vnabla{}^\mu, \vnabla{}^\nu] &= 0, \\
    \hnabla_\mu p_\alpha = \hnabla_\mu g_{\alpha \beta} = \vnabla{}^\mu g_{\alpha \beta} &= 0.    
\end{split}
\end{equation}

\section{Variation of the action}
\label{app:ELE}

Here, we derive the Euler-Lagrange equations that correspond to the action
\begin{equation}
    J = \int_M \mathrm{d}^4 x \,  \sqrt{g} \, \mathcal{L},
\end{equation}
where the Lagrangian density is of the following form:
\begin{align}
\begin{split}
\mathcal{L} = \mathcal{L} \Big( S&(x), \nabla_\mu S (x), \\ &A_\alpha [x, \nabla S(x)], \nabla_\mu \left\{A_\alpha [x, \nabla S(x)] \right\}, \\ &A^{*\alpha} [x, \nabla S(x)], \nabla_\mu \left\{ A^{*\alpha} [x, \nabla S(x)] \right\} \Big).
\end{split}
\end{align}
Here, $S(x)$ is an independent field, while $A_\alpha$ and $ A^{*\alpha}$ cannot be considered independent, since they depend on $\nabla_\mu S$. Following Hawking and Ellis \cite[p. 65]{hawking1973}, we define the variation of a field $\Psi_i$ as a one-parameter family of fields $\Psi_i (u, x)$, with $u \in (-\varepsilon, \varepsilon)$ and $x \in M$. We use the following notation:
\begin{equation}
    \frac{\partial \Psi_i(u,x)}{\partial u} \Bigg|_{u=0} = \Delta \Psi_i.
\end{equation}
Note that the derivative with respect to the parameter $u$ commutes with the covariant derivative, so we have:
\begin{widetext}
\begin{align}
    \frac{\mathrm{d}}{\mathrm{d}u} \nabla_\mu S(u, x) &= \nabla_\mu \left( \frac{\partial S}{\partial u} \right) , \\
    \frac{\mathrm{d}}{\mathrm{d} u} A_\alpha \left(u, x, \nabla S (u, x) \right) &= \frac{\partial A_\alpha}{\partial u} + \frac{\partial A_\alpha}{\partial \nabla_\nu S} \nabla_\nu \left( \frac{\partial S}{\partial u} \right) , \\
    \begin{split}
    \frac{\mathrm{d}}{\mathrm{d} u} \nabla_\mu \left[ A_\alpha \left(u, x, \nabla S(u, x) \right) \right] &= \nabla_\mu \left[ \frac{\mathrm{d}}{\mathrm{d} u} A_\alpha \left(u, x, \nabla S(u, x) \right) \right] = \nabla_\mu \left[ \frac{\partial A_\alpha}{ \partial u} + \frac{\partial A_\alpha}{\partial \nabla_\nu S} \nabla_\nu \left( \frac{\partial S}{\partial u} \right) \right].
    \end{split}
\end{align}
We consider the variation of the action, taking special care when applying the chain rule:
\begin{equation}
\begin{split}
    0 = \frac{\mathrm{d} J}{\mathrm{d} u} \Bigg|_{u=0} = \int_M \mathrm{d}^4 x \,  \sqrt{g} \, \Bigg\{& \frac{\partial \mathcal{L}}{\partial S} \Delta S + \frac{\partial \mathcal{L}}{\partial \nabla_\mu S} \Delta \left( \nabla_\mu S \right) \\
    &+ \frac{\partial \mathcal{L}}{\partial A_\alpha} \left[ \Delta A_\alpha + \frac{\partial A_\alpha}{\partial \nabla_\mu S}  \nabla_\mu \left( \Delta S \right) \right] + \frac{\partial \mathcal{L}}{\partial \nabla_\mu A_\alpha} \nabla_\mu \left[ \Delta A_\alpha  + \frac{\partial A_\alpha}{\partial \nabla_\nu S} \nabla_\nu \left( \Delta S \right) \right]\\
    &+ \frac{\partial \mathcal{L}}{\partial A^{*\alpha}} \left[ \Delta A^{*\alpha} + \frac{\partial A^{*\alpha}}{\partial \nabla_\mu S} \nabla_\mu \left( \Delta S \right) \right] + \frac{\partial \mathcal{L}}{\partial \nabla_\mu A^{*\alpha}} \nabla_\mu \left[ \Delta  A^{*\alpha}  + \frac{\partial A^{*\alpha}}{\partial \nabla_\nu S} \nabla_\nu \left( \Delta S \right) \right] \Bigg\}.
\end{split}
\end{equation}
Integrating by parts and assuming the boundary terms vanish, we obtain:
\begin{equation}
\begin{split}
    0 = \frac{\mathrm{d} J}{\mathrm{d} u}& \Bigg|_{u=0} = \int_M \mathrm{d}^4 x \,  \sqrt{g} \, \Bigg\{ \left(\frac{\partial \mathcal{L}}{\partial A_\alpha} - \nabla_\mu \frac{\partial \mathcal{L}}{\partial \nabla_\mu A_\alpha} \right) \Delta A_\alpha + \left(\frac{\partial \mathcal{L}}{\partial A^{*\alpha}} - \nabla_\mu \frac{\partial \mathcal{L}}{\partial \nabla_\mu A^{*\alpha} } \right) \Delta A^{*\alpha}  \\
     &+ \frac{\partial \mathcal{L}}{\partial S} \Delta S - \nabla_\mu \Bigg[ \frac{\partial \mathcal{L}}{\partial \nabla_\mu S} + \frac{\partial A_\alpha}{\partial \nabla_\mu S} \left(\frac{\partial \mathcal{L}}{\partial A_\alpha} - \nabla_\nu  \frac{\partial \mathcal{L}}{\partial \nabla_\nu A_\alpha} \right)   + \frac{\partial A^{*\alpha}}{\partial \nabla_\mu S} \left(\frac{\partial \mathcal{L}}{\partial A^{*\alpha}} - \nabla_\nu \frac{\partial \mathcal{L}}{\partial \nabla_\nu A^{*\alpha}} \right) \Bigg]\Delta S \Bigg\}.
\end{split}
\end{equation}
Since the above equation must be satisfied for all variations $\Delta S$, $\Delta A_\alpha$, and $\Delta A^{*\alpha}$, we obtain the following Euler-Lagrange equations:
\begin{align}
    \frac{\partial \mathcal{L}}{\partial A^{*\alpha}} - \nabla_\mu \frac{\partial \mathcal{L}}{\partial \nabla_\mu A^{*\alpha}} &= \mathcal{O}(\epsilon^2) , \label{eq:ELE1} \\
    \frac{\partial \mathcal{L}}{\partial A_{\alpha}} - \nabla_\mu \frac{\partial \mathcal{L}}{\partial \nabla_\mu A_{\alpha}} &= \mathcal{O}(\epsilon^2) \label{eq:ELE2},\\
    \begin{split}
    \frac{\partial \mathcal{L}}{\partial S} - \nabla_\mu   \Bigg[ \frac{\partial \mathcal{L}}{\partial \nabla_\mu S} + \frac{\partial A_\alpha}{\partial \nabla_\mu S} \left(\frac{\partial \mathcal{L}}{\partial A_\alpha} - \nabla_\nu  \frac{\partial \mathcal{L}}{\partial \nabla_\nu A_\alpha} \right) + \frac{\partial A^{*\alpha}}{\partial \nabla_\mu S} \left(\frac{\partial \mathcal{L}}{\partial A^{*\alpha}} - \nabla_\nu \frac{\partial \mathcal{L}}{\partial \nabla_\nu A^{*\alpha}} \right) \Bigg] &= \mathcal{O}(\epsilon^2).
    \end{split} \label{eq:ELE3}
\end{align}
\end{widetext}
Furthermore, Eq.~\eqref{eq:ELE3} can be simplified by using Eqs.~\eqref{eq:ELE1} and \eqref{eq:ELE2}. Thus, as a final result, we have the following set of Euler-Lagrange equations:
\begin{equation} \label{eq:ELE}
\begin{split}
    \frac{\partial \mathcal{L}}{\partial A^{*\alpha}} - \nabla_\mu \frac{\partial \mathcal{L}}{\partial \nabla_\mu A^{*\alpha}} &= \mathcal{O}(\epsilon^2) , \\
    \frac{\partial \mathcal{L}}{\partial A_{\alpha}} - \nabla_\mu \frac{\partial \mathcal{L}}{\partial \nabla_\mu A_{\alpha}} &= \mathcal{O}(\epsilon^2) ,\\
    \frac{\partial \mathcal{L}}{\partial S} - \nabla_\mu \frac{\partial \mathcal{L}}{\partial \nabla_\mu S} &= \mathcal{O}(\epsilon^2).
\end{split}
\end{equation}

\section{Berry curvature} \label{app:Berry_curvature}

In order to calculate the Berry curvature terms \eqref{eq:Berry_curvature}, it is enough to use a tetrad $\{ t^\alpha, p^\alpha, v^\alpha, w^\alpha \}$, where $t^\alpha$ is a future-oriented timelike vector field representing a family of observers and $p^\alpha$ is a generic vector, not necessarily null, representing the momentum of a point particle (ray). The vectors $v^\alpha$ and $w^\alpha$ are real spacelike vectors related to $m^\alpha$ and $\bar{m}^\alpha$ by the following relations:
\begin{equation} \label{eq:circ-to-linear}
    m^\alpha = \frac{1}{\sqrt{2}} \left( v^\alpha + i w^\alpha \right), \qquad \bar{m}^\alpha = \frac{1}{\sqrt{2}} \left( v^\alpha - i w^\alpha \right).
\end{equation}
The elements of the tetrad $\{ t^\alpha, p^\alpha, v^\alpha, w^\alpha \}$ satisfy the following relations:
\begin{equation} \label{eq:tetrad_prop}
\begin{gathered}
    t_\alpha t^\alpha = -1, \qquad p_\alpha p^\alpha = \kappa, \qquad t_\alpha p^\alpha = -\epsilon \omega, \\
    v_\alpha v^\alpha = w_\alpha w^\alpha = 1,\\  
    t_\alpha v^\alpha = t_\alpha w_\alpha = p_\alpha v^\alpha = p_\alpha w^\alpha = v_\alpha w^\alpha = 0.
\end{gathered}
\end{equation}
Note that the vectors $v^\alpha$ and $w^\alpha$ depend of $p^\mu$ through the orthogonality condition, while $t^\alpha$ is independent of $p^\mu$. We start by computing the vertical derivatives of the vectors $v^\alpha$ and $w^\alpha$. Using the tetrad, we can write:
\begin{align}
    \vnabla{}^\mu v^\alpha = \frac{\partial v^\alpha}{ \partial p_\mu} &= {c_1}^\mu t^\alpha + {c_2}^\mu p^\alpha + {c_3}^\mu v^\alpha + {c_4}^\mu w^\alpha, \\
    \vnabla{}^\mu w^\alpha = \frac{\partial w^\alpha}{ \partial p_\mu} &= {d_1}^\mu t^\alpha + {d_2}^\mu p^\alpha + {d_3}^\mu v^\alpha + {d_4}^\mu w^\alpha,
\end{align}
where ${c_i}^\mu$ and ${d_i}^\mu$ are unknown vector fields that need to be determined. Using the properties from Eq.~\eqref{eq:tetrad_prop}, we obtain
\begin{equation} \label{eq:vert_deriv}
\begin{split}
     \vnabla{}^\mu v^\alpha =  \frac{\epsilon \omega}{\epsilon^2 \omega^2 + \kappa} v^\mu t^\alpha - \frac{1}{\epsilon^2 \omega^2 + \kappa} v^\mu p^\alpha + {c_4}^\mu w^\alpha, \\
     \vnabla{}^\mu w^\alpha =  \frac{\epsilon \omega}{\epsilon^2 \omega^2 + \kappa} w^\mu t^\alpha - \frac{1}{\epsilon^2 \omega^2 + \kappa} w^\mu p^\alpha + {d_3}^\mu v^\alpha.
\end{split}
\end{equation}
Applying the same arguments to the terms $\nabla_\mu v_\alpha$ and $\nabla_\mu w_\alpha$, we also obtain
\begin{equation} \label{eq:cov_deriv}
\begin{split}
    &\begin{split} \nabla_\mu v_\alpha =  &-\frac{1}{\epsilon^2 \omega^2 + \kappa} \left( \epsilon \omega p_\sigma \nabla_\mu v^\sigma + \kappa t_\sigma \nabla_\mu v^\sigma \right) t_\alpha \\  &+\frac{1}{\epsilon^2 \omega^2 + \kappa} \left( p_\sigma \nabla_\mu v^\sigma - \epsilon \omega t_\sigma \nabla_\mu v^\sigma \right) p_\alpha + {f_4}_\mu w_\alpha, \end{split} \\
    &\begin{split} \nabla_\mu w_\alpha =  &-\frac{1}{\epsilon^2 \omega^2 + \kappa} \left( \epsilon \omega p_\sigma \nabla_\mu w^\sigma + \kappa t_\sigma \nabla_\mu w^\sigma \right) t_\alpha \\  &+\frac{1}{\epsilon^2 \omega^2 + \kappa} \left( p_\sigma \nabla_\mu w^\sigma - \epsilon \omega t_\sigma \nabla_\mu w^\sigma \right) p_\alpha + {g_3}_\mu v_\alpha. \end{split}
\end{split}
\end{equation}
Note that the fields ${c_4}_\mu$, ${d_3}_\mu$, ${f_4}_\mu$, and ${g_3}_\mu$ are undetermined within this approach, but this is not a problem, because they do not affect the Berry curvature.

\subsection{$\boldsymbol{F_{p p}}$}

We compute $\left({F_{p p}}\right)^{\nu \mu}$ by using Eq.~\eqref{eq:vert_deriv} and setting $\kappa = 0$. Since vertical derivatives commute (see Eq.~\eqref{eq:hor_vert_prop}), we can write
\begin{equation}
\begin{split}
    \left({F_{p p}}\right)^{\nu \mu} &= i \left( \vnabla{}^\mu \bar{m}^\alpha \vnabla{}^\nu m_\alpha - \vnabla{}^\nu \bar{m}^\alpha \vnabla{}^\mu m_\alpha \right) \\
    &= \vnabla{}^\nu v^\alpha \vnabla{}^\mu w_\alpha - \vnabla{}^\mu v^\alpha \vnabla{}^\nu w_\alpha \\
    &= \frac{2}{\epsilon^2 \omega^2} v^{[\nu}w^{\mu]} \\
    &= \frac{2i}{\epsilon^2 \omega^2} m^{[\nu} \bar{m}^{\mu]}.
\end{split}
\end{equation}

\subsection{$\boldsymbol{F_{x x}}$}

We have
\begin{equation}
\begin{split}
     \left({F_{xx}}\right)_{\nu \mu} = i \Big(& \nabla_\mu \bar{m}^\alpha  \nabla_\nu m_\alpha - \nabla_\nu \bar{m}^\alpha \nabla_\mu m_\alpha \\ &+ \bar{m}^\alpha \nabla_{[\mu} \nabla_{\nu]} m_\alpha - m_\alpha \nabla_{[\mu} \nabla_{\nu]} \bar{m}^\alpha \Big). 
\end{split}
\end{equation}
The last two terms can be expressed in terms of the Riemann tensor:
\begin{equation}
\begin{split}
    i \Big( \bar{m}^\alpha \nabla_{[\mu} \nabla_{\nu]} m_\alpha - m_\alpha \nabla_{[\mu} \nabla_{\nu]} \bar{m}^\alpha \Big) = - i R_{\alpha \beta \mu \nu} m^\alpha \bar{m}^\beta.
\end{split}
\end{equation}
The first two terms can be computed using Eq.~\eqref{eq:cov_deriv} and $\kappa=0$:
\begin{equation}
\begin{split}
  (\tilde{F}_{x x})_{\nu \mu} &= i \Big( \nabla_\mu \bar{m}^\alpha \nabla_\nu m_\alpha - \nabla_\nu \bar{m}^\alpha \nabla_\mu m_\alpha \Big) \\
  &= \nabla_\nu v^\alpha \nabla_\mu w_\alpha - \nabla_\mu v^\alpha \nabla_\nu w_\alpha \\
   &\begin{split}=&\frac{1}{\epsilon^2 \omega^2} \Big( p_\sigma \nabla_\mu v^\sigma p_\rho \nabla_\nu w^\rho - p_\sigma \nabla_\nu v^\sigma p_\rho \nabla_\mu w^\rho \\
   &- \epsilon \omega p_\sigma \nabla_\mu v^\sigma t_\rho \nabla_\nu w^\rho + \epsilon \omega p_\sigma \nabla_\nu v^\sigma t_\rho \nabla_\mu w^\rho \\
   &- \epsilon \omega t_\sigma \nabla_\mu v^\sigma p_\rho \nabla_\nu w^\rho + \epsilon \omega t_\sigma \nabla_\nu v^\sigma p_\rho \nabla_\mu w^\rho \Big)
   \end{split} \\
    &\begin{split}=&\frac{1}{\epsilon^2 \omega^2} \Big( p_\sigma \nabla_\mu m^\sigma p_\rho \nabla_\nu \bar{m}^\rho - p_\sigma \nabla_\nu m^\sigma p_\rho \nabla_\mu \bar{m}^\rho \\
   &- \epsilon \omega p_\sigma \nabla_\mu m^\sigma t_\rho \nabla_\nu \bar{m}^\rho + \epsilon \omega p_\sigma \nabla_\nu m^\sigma t_\rho \nabla_\mu \bar{m}^\rho \\
   &- \epsilon \omega t_\sigma \nabla_\mu m^\sigma p_\rho \nabla_\nu \bar{m}^\rho + \epsilon \omega t_\sigma \nabla_\nu m^\sigma p_\rho \nabla_\mu \bar{m}^\rho \Big).
   \end{split}
\end{split}
\end{equation}

\subsection{$\boldsymbol{F_{p x}}$ and $\boldsymbol{F_{x p}}$}

Since $\left({F_{p x}}\right)\indices{_\nu^\mu} = - \left({F_{x p}}\right)\indices{^\mu_\nu}$, it is enough to compute only one term. Using Eqs.~\eqref{eq:vert_deriv} and \eqref{eq:cov_deriv}, and setting $\kappa=0$, we obtain
\begin{equation}
\begin{split}
    \left({F_{p x}}\right)\indices{_\nu^\mu} &= i \left( \vnabla{}^\mu \bar{m}^\alpha \nabla_\nu m_\alpha - \nabla_\nu \bar{m}^\alpha \vnabla{}^\mu m_\alpha \right) \\
    &= \nabla_\nu v^\alpha \vnabla{}^\mu w_\alpha - \vnabla{}^\mu v^\alpha \nabla_\nu w_\alpha \\
    &= \frac{1}{\epsilon^2 \omega^2} \Big[ \left( p_\sigma \nabla_\nu w^\sigma - \epsilon \omega t_\sigma \nabla_\nu w^\sigma \right) v^\mu \\
    &\qquad\quad- \left( p_\sigma \nabla_\nu v^\sigma - \epsilon \omega t_\sigma \nabla_\nu v^\sigma \right) w^\mu \Big] \\
    &= \frac{i}{\epsilon^2 \omega^2} \Big[ \left( p_\sigma \nabla_\nu \bar{m}^\sigma - \epsilon \omega t_\sigma \nabla_\nu \bar{m}^\sigma \right) m^\mu \\
    &\qquad\quad- \left( p_\sigma \nabla_\nu m^\sigma - \epsilon \omega t_\sigma \nabla_\nu m^\sigma \right) \bar{m}^\mu  \Big].
\end{split}
\end{equation}

\section{Coordinate transformation} \label{app:coord}

The substitution from Eqs.~\eqref{eq:coord} and \eqref{eq:coord1} can be obtained, up to terms of order $\epsilon^2$, as a linearization of the following composition of changes of coordinates on the cotangent bundle $T^* M$. Consider the family of diffeomorphisms $(\Phi_\epsilon)$ generated by the vector field on $M$ 
\begin{equation}
Y = is \bar{m}^{\alpha} \vnabla{}^\mu {m}_\alpha \partial_{x^\mu},
\end{equation}
that is to say
\begin{equation}
\dfrac{\text{d}}{\text{d} \epsilon }\Phi_\epsilon (x) = Y(\Phi_\epsilon (x)) \text{ with }\Phi_0 (x)  =x.
\end{equation}
By construction, the Taylor expansion in a coordinate chart of $\Phi_\epsilon$ at order $\epsilon^1$ leads to Eq.~\eqref{eq:coord}. $\Phi_\epsilon$ naturally lifts to the cotangent bundle using the pullback $\Phi_\epsilon^\ast$:  
\begin{equation}
    \Phi_\epsilon^\ast: (x, p) \mapsto (\Phi_\epsilon(x), p \circ d\Phi^{-1}_\epsilon|_{\Phi_\epsilon (x)}).
\end{equation}
Note that the choice of the lift is not unique. The mapping $ \Phi_\epsilon^\ast$ is, at order one in $\epsilon$, in coordinates,
\begin{equation}
(x^\mu,p_\mu) \mapsto (x^\mu + is\epsilon \bar{m}^{\alpha} \vnabla{}^\mu {m}_\alpha , p_\mu - i\epsilon s p_\beta  \partial_{x^\mu} ( \bar{m}^{\alpha} \vnabla{}^\beta {m}_\alpha) ). 
\end{equation}
Consider next the translation of the momentum variable defined by 
\begin{equation}
\Psi_\epsilon: (x, p) \mapsto (x, p -\epsilon \sigma ), 
\end{equation}
where $\sigma =  is( \bar{m}^{\alpha} \nabla_\mu {m}_\alpha + p_\beta  \partial_{x^\mu} ( \bar{m}^{\alpha} \vnabla{}^\beta {m}_\alpha))dx^\mu$.
The linearization in $\epsilon$ of the diffeomorphism $\Psi_\epsilon\circ {\Phi_\epsilon}^\ast$ provides by construction the change of variables in Eqs.~\eqref{eq:coord} and \eqref{eq:coord1}.

\bibliography{references}

\end{document}